\documentclass[12pt]{article}
\usepackage{amssymb, amsfonts, graphicx, hyperref, psfrag, epsfig}
\newcommand{\be}{\begin{equation}}
\newcommand{\ee}{\end{equation}}
\newcommand{\bea}{\begin{eqnarray}}
\newcommand{\eea}{\end{eqnarray}}
\newcommand{\ba}{\begin{array}}
\newcommand{\ea}{\end{array}}
\newcommand{\pt}{\partial}
\newcommand{\nn}{\nonumber}

\csname @addtoreset\endcsname{equation}{section}

\begin{document}
\begin{titlepage}

\title{\bf\Large  Holographic model with a NS-NS field \vspace{18pt}}

\author{\normalsize Yunseok Seo$^a$, Sang-jin Sin$^b$ and Wei-shui Xu$^{a,b}$ \vspace{12pt}\\
${}^a${\it\small Center for Quantum Spacetime, Sogang University, Seoul 121-742, Korea}\\ 
${}^b${\it\small Physics Department, Hanyang University, Seoul 133-791, Korea}\\
{\small e-mail: { \it yseo, sjsin@hanyang.ac.kr,~\it wsxuitp@gmail.com}}
}

\date{}
\maketitle \voffset -.2in \vskip 1cm 
\centerline{\bf Abstract} \vskip .4cm 

We consider a holographic model constructed through using the
D4/D8-$\overline{\rm D8}$ brane configuration with a NS-NS background
field. We study some properties of the effective field theory in this
intersecting brane construction, and calculate the effects of this NS-NS
background field on some underlying dynamics. We also discuss some other general
brane configurations.

\vskip 7.5cm \noindent June 2009 
\thispagestyle{empty}

\end{titlepage}

\newpage

\section{Introduction}
The AdS/CFT correspondence proposed by Maldacena is one realization of the
holographic principle \cite{'tHooft:1993gx}. It means
IIB string theory on the $AdS_5\times S^5$ gravity background is dual to the
$\mathcal{N}=4$ supersymmetric Yang-Mills field theory in the boundary
\cite{Maldacena:1997re}-\cite{Aharony:1999ti}. In order to use the AdS/CFT
method to study more realistic physics, one need to consider some necessary
properties, such as less supersymmetries, confinement and chiral symmetry
breaking in constructing models. It was firstly given some investigations in the
references \cite{Witten:1998qj} and
\cite{Polchinski:2000uf}. In \cite{Karch:2002sh}, Karch
et al. introduce fundamental flavors into the holographic models by adding some
probe flavor D-branes. In the quenching approximation, the supergravity method
is a useful way to study some strong coupling dynamics of the boundary field
theory. Follow this proposal, many holographic models are constructed to study
the strong coupling physics by the supergravity approximation
\cite{Kruczenski:2003be}-\cite{Antonyan:2006vw}. If the number of flavor
branes is almost equal to the color brane number ($N_f\sim N_c$), it means the
flavor backreaction becomes large, then the quenching approximation isn't
reliable \cite{Burrington:2004id}.

Now we generalize to consider the color brane background with a NS-NS $B_{MN}$
background field. In string theory, we know that a NS-NS $B_{MN}$ background
field produces some non-commutative effects in the field theory on the D-brane
world-volume \cite{Seiberg:1999vs}, \cite{Maldacena:1999mh} and
\cite{Alishahiha:1999ci}. The strong and weak coupled regimes of these
non-commutative effective field theory can be analyzed as the same way in
\cite{Itzhaki:1998dd}. After adding the flavor D-brane into such background, one
can obtain some non-commutative QCD-like effective field theory on the
intersecting region of the brane configuration. Such model was firstly studied
in \cite{Arean:2005ar}. 
It is interesting to construct some holographic models with a NS-NS
$B_{MN}$ field from string theory, and investigate some properties of these
models by using the gauge/gravity correspondence. Then we can understand such
NS-NS $B_{MN}$ background field how to affect the underlying strong coupled
dynamics of the effective field theory.

In this paper, we construct a holographic model through using the brane
configuration D4/D8-$\overline{\rm D8}$ just like the Sakai-Sugimoto model
\cite{Sakai:2004cn}. The difference with the Sakai-Sugimoto model is now the
color D4-branes gravity background with a NS-NS $B_{12}$ field (see appendix). If
there only exists a constant NS-NS $B_{12}$ background, then it will be
equivalent to turn on a magnetic field on the flavor brane worldvolume by the
gauge invariance. But now except for this NS-NS field, the dilaton and R-R field
also include this non-commutative parameter $b$. Thus, it can't be gauged away
by add an additional gauge field on the flavor brane.

Then the effective field theory on the intersecting region of this brane
configuration is four-dimensional and non-commutative along the coordinates
$x_1$ and $x_2$. And some strong coupled physics of this effective theory can be
studied through investigating the dynamics of the flavor D8-brane in the
D4-brane gravity background. If one don't consider the gauge field on the flavor
D8-brane, then the effective D8-brane action is same as the commutative case in
\cite{Sakai:2004cn}. The reason is the non-commutative parameter from the
dilaton will cancel out with the contribution of the square root part in the
D8-brane DBI action. It means that this NS-NS $B_{12}$ background field doesn't
have any effects on the dynamics of the flavor D8-brane. But after a magnetic
field along the $x_1$ and $x_2$ directions is turned on, then the
non-commutative parameter will combine together with this magnetic field, and
can't be canceled in the effective D8-brane action. Finally, this effective DBI
action contains the contribution of the NS-NS $B_{12}$ field. And its effects on
the chiral symmetry breaking and phase transition etc. can be analyzed in
details through using the supergravity/Born-Infeld approximation. We also
investigate a fundamental string moving in the near horizon geometry of the
black D4-branes. And we calculate the drag force of moving quark and the Regge
trajectory of mesons in this non-commutative quark-gluon-plasma(QGP), and
analyze the NS-NS $B_{12}$ how to influence on these quantities.
 
This papers is organized as follows. In the section 2, we firstly give a brane
construction, and study the D8-brane dynamics without and with a magnetic field
in the near horizon geometry of the color D4-branes background with a NS-NS
$B_{12}$ field. In the section 3, we generalize to study some other brane
configurations by using the same way as in the previous section. Then in the
section 4, we study the physics of a fundamental string moving through the near
horizon background of the black D4-branes with a NS-NS $B_{12}$ field. The
section 5 is some conclusions and discussions. And finally there is an appendix.

\section{Brane configuration}
We consider a brane configuration which is composed of D4, D8 and $\overline{\rm
  D8}$ brane in IIA string theory. The $N_c$ D4 are the color branes , while the D8 and
$\overline{\rm D8}$ branes produce the flavor degrees of freedom. Their
extending directions of these branes are as follows 
\be
\begin{array}{ccccccccccccc}
 &0 &1 &2 &3 &4 &5 &6 &7 &8 &9\\
N_c~{\rm D4}: &{\rm x} &{\rm x} &{\rm x} &{\rm x} &{\rm x} &{} &{} &{} &{} &{}\\
N_f~{\rm D8,\overline{D8}}: &{\rm x} &{\rm x} &{\rm x} &{\rm x} &{} &{\rm x}
&{\rm x} &{\rm x} &{\rm x}  &{\rm  x}
\end{array}
\label{configuration} \ee 
In this brane configuration, the D8 and $\overline{\rm
  D8}$ branes are parallel each other and intersect with the $N_c$ D4 branes
along directions $(x_0,\cdots, x_3)$. All the others are the transverse
directions to the intersection region. We assume the coordinate $x_4$ to be
compactified on a circle $S^1$ , so it satisfies a periodic condition $x_4\sim
x_4+ \delta x_4$. Then the fermions on the D4 brane with anti-periodic boundary
condition on this circle get mass and are decoupled from the low energy
effective theory. Also we assume the number of the color and flavor branes
satisfies the condition $N_f\ll N_c$. In this quenching approximation, the
backreaction of the flavor branes on the color branes can be ignored. Due to the
existence of the NS-NS $B_{12}$ field, the coordinates $x_1$ and $x_2$ are
non-commutative \be [x_1,~x_2]\sim b,\ee the low energy effective theory in the
intersecting region is a non-commutative field theory. If let the
non-commutative parameter vanish, the gravity background here reduces to the
usual near horizon geometry of D4 branes without a NS-NS background field. So
our model correspondingly goes to the Sakai-Sugimoto model
\cite{Sakai:2004cn}. This means that this holographic model is connected to
Sakai-Sugimoto model through varying the non-commutative parameter $b$. Thus,
the holographic model with a NS-NS $B_{12}$ background field can be regarded as
a non-commutative deformation to the Sakai-Sugimoto holographic model.

After compactifying the coordinate $x_4$ and using the gravity background
(\ref{metric2}) introduced in the appendix, we get the following gravity solution
\bea &&
ds^2=\left(\frac{u}{R}\right)^{3/2}\left[-dt^2+h(dx_1^2+dx_2^2)+dx_3^2+fdx_4^2\right]
+\left(\frac{R}{u} \right)^{3/2}(\frac{du^2}{f}+u^2d\Omega_4^2),\cr
&&R^3=\pi^4g_sN_c, \quad h=\frac{1}{1+a^3u^3},\quad
a^3\equiv\frac{b^2}{R^3},\quad H=1-\frac{u_{KK}^3}{u^3}, \nn \\
&& B_{12}=\frac{a^3u^3}{b(1+a^3u^3)}=\frac{1}{b}(1-h), \quad
e^{2\phi}=\frac{g_s^2u^{3/2}}{R^{3/2}b^2}h,  \label{metric5}\\
&& C_{01234}=g_s^{-1}h,\quad C_{034}=g_s^{-1}b.\nn \eea In order to avoid the
singularity, the coordinate $x_4$ will be periodic with a radius \be x_4\sim
x_4+ \delta x_4=x_4+\frac{4\pi}{3}\frac{R^{3/2}}{u_{KK}^{1/2}}. \ee From this
radius of the coordinate $x_4$, we know it
corresponds to a KK mass scale \be M_{KK}=\frac{2\pi}{\delta
  x_4}=\frac{3}{2}\frac{u_{KK}^{1/2}}{R^{3/2}}.\ee

By using the solution (\ref{metric5}), and doing a double Wick rotation between the
coordinate $t$ and $x_4$, we obtain a black hole solution as follows \bea &&
ds^2=\left(\frac{u}{R}\right)^{3/2}\left[-Hdt^2+h(dx_1^2+dx_2^2)+dx_3^2+dx_4^2\right]
+\left(\frac{R}{u} \right)^{3/2}(\frac{du^2}{H}+u^2d\Omega_4^2),\cr
&&R^3=\pi^4g_sN_c, \quad h=\frac{1}{1+a^3u^3},\quad
a^3\equiv\frac{b^2}{R^3},\quad H=1-\frac{u_H^3}{u^3}, \nn \\
&& B_{12}=\frac{a^3u^3}{b(1+a^3u^3)}=\frac{1}{b}(1-h), \quad
e^{2\phi}=\frac{g_s^2u^{3/2}}{R^{3/2}b^2}h,  \label{metric6}\\
&& C_{01234}=g_s^{-1}h,\quad C_{034}=g_s^{-1}b.\nn \eea Now the coordinates $t$
and $x_4$ are all compactified. The time coordinate $t$ need to be periodic \be
t\sim t+\delta t=t+\frac{4\pi}{3}\frac{R^{3/2}}{u_H^{1/2}},\ee then this
background is well-defined. And the corresponding temperature of this black hole
is \be T=\frac{3}{4\pi}\frac{u_H^{1/2}}{R^{3/2}}.\ee Through a Hawking-Page
phase transition at a critical point $u_H=u_{KK}$, the low temperature solution
(\ref{metric5}) will go to the black hole background (\ref{metric6}).

\subsection{Low temperature phase}
For the low temperature phase, the dominated gravity background is the equation
(\ref{metric5}). Now we consider the dynamics of the probe D8-${\rm
  \overline{D8}}$ brane in this background. Assume the transverse coordinate
$x_4$ of D8 branes is a function of the radial coordinate $u$,
i.e. $x_4=x_4(u)$. Then the induced metric on the D8 branes is \bea
ds^2&=&\left(\frac{u}{R}\right)^{3/2}\left[-dt^2+h(dx_1^2+dx_2^2)+dx_3^2\right]+\left(\frac{R}{u}
\right)^{3/2}u^2d\Omega_4^2 \cr && \quad
+\left[\left(\frac{R}{u}\right)^{3/2}f^{-1}+f\left(\frac{u}{R}\right)^{3/2}\left(\frac{\pt
      x_4}{\pt u}\right)^2\right]du^2, \label{inducemetric1}\eea

Using the D-brane effective action \bea && S=S_{DBI}+S_{CS}, \cr
&&S_{DBI}=-T_p\int d^{p+1}\xi e^{-\phi}\sqrt{-\det(P[G+B]_{\mu\nu}+2\pi
  F_{\mu\nu})},\\
&& S_{CS}=T_p\int\sum_iP[C_i\wedge e^{B_2}]\wedge e^{2\pi F_2}, \nn \eea then we
get the D8-brane action is {\footnote{Here we don't turn on the gauge field on
    the D8 brane worldvolume, so to investigate the dynamics of D8-brane in the
    gravity background (\ref{metric5}), it is enough to only use the effective
    action of one single flavor D8 brane.}} \bea S_{DBI}\sim\int du
u^{13/4}\sqrt{
  \left(\frac{R}{u}\right)^{3/2}f^{-1}+f\left(\frac{u}{R}\right)^{3/2}\left(\frac{\pt
      x_4}{\pt u}\right)^2}, \label{action1}\eea and \be S_{CS}=0. \ee Except
for the proportional coefficient, this action
is same as the D8-brane effective in the D4-branes background without a
NS-NS $B_{12}$ background field. The reason is the $h$ included in the term
$e^{-\phi}$ will be canceled out with the same factor in the square
root part of the DBI action. Thus, at low temperature, the NS-NS background field doesn't influence on
the dynamics of the probe D8-brane in this non-commutative gravity
background. Its dynamics is same with the usual commutative case
\cite{Sakai:2004cn} and \cite{Aharony:2006da}. 

Then one can get two different solutions from the equation of motion: one is the
connected configuration of the D8-${\rm \overline{D8}}$ branes, the other is the
separated case. The connected solution corresponds to the chiral symmetry
breaking phase, while the chiral symmetry is preserved for the separated
solution. But the on-shell energy difference between the connected and separated
solution is always negative, so this connected configuration is dominated in the
low temperature phase \cite{Aharony:2006da}. The
chiral symmetry is always broken $U(N_f)\times U(N_f)\rightarrow U(N_f)_{\rm
  diag}$, and the corresponding Nambu-Goldstone boson can be found through
calculating the meson spectra \cite{Sakai:2004cn}.

\subsection{High temperature phase}
In the high temperature phase, the corresponding gravity background is the black
solution (\ref{metric6}). We still assume the transverse coordinate $x_4$ of the
flavor D8-brane depends on the coordinate $u$, then the induced metric on the
D8-brane is \bea ds^2&=&
\left(\frac{u}{R}\right)^{3/2}\left[-Hdt^2+h(dx_1^2+dx_2^2)+dx_3^2\right]\cr
&&\quad \quad
+\left(\left(\frac{R}{u}\right)^{3/2}H^{-1}+\left(\frac{u}{R}\right)^{3/2}\left(\frac{\pt
      x_4}{\pt u}\right)^2\right)du^2
+\left(\frac{R}{u}\right)^{3/2}u^2d\Omega_4^2. \label{inducemetric2}\eea So the
effective action of the D8-brane is \be S\sim \int du
u^{13/4}H^{1/2}\sqrt{\left(\frac{R}{u}\right)^{3/2}H^{-1}+\left(\frac{u}{R}\right)^{3/2}
  \left(\frac{\pt x_4}{\pt u}\right)^2}.\ee Clearly, it still doesn't depend on
the NS-NS background field, and is also same as the commutative case
\cite{Sakai:2004cn} except for the proportional coefficient. 
Then all the other analysis of the D8-${\rm
  \overline{D8}}$ dynamics will be same. Through comparing the energy of these
two solutions, one can find there exists two phase at high temperature. One is
the chiral symmetry breaking phase, the other is chiral symmetry restoration
phase. Between these two phases, it has a chiral phase transition at a critical
temperature. Below this critical temperature the system is located at the chiral
symmetry breaking phase i.e. $U(N_f)\times U(N_f)\rightarrow U(N_f)_{\rm diag}$,
otherwise it will be in the chiral symmetry restoration phase.

\subsection{Adding a magnetic field}
Now we introduce a constant magnetic field along the $x_1$ and $x_2$ directions
\be 2\pi F_{12}=B \label{magnetic},\ee on the flavor
D8-brane\footnote{ Holographic models with a constant magnetic or electric field were studied in
  \cite{Filev:2007gb}-\cite{Johnson:2009ev}.}. This magnetic field is equivalent
to a constant NS-NS $B'_{12}$ field by the gauge invariance on the
D8-brane. This NS-NS $B'_{12}$ field will contribute a constant transportation
to the NS-NS $B_{12}$ field \be B_{\rm
  new}=B_{12}+B'_{12}=B+\frac{1-h(u)}{b}. \label{NSNS}\ee This new background is still a
solution of the IIA supergravity. Since the dilation and R-R field in the
gravity background all depend on the parameter $b$, this magnetic field $B$
can't be canceled out through a redefinition of the parameter $b$. Then the
flavor D8-brane effective action contains the contributions of the NS-NS
$B_{12}$ and magnetic field together. In the following we will analyze the
effect of the parameters $b$ and $B$ on the flavor brane dynamics in details.

\subsubsection{Low temperature}
The induced metric on the flavor D8-brane is same as the equation
(\ref{inducemetric1}). After a magnetic field (\ref{magnetic}) is turned on, the
effective action of the probe D8 brane is \bea S &\sim& \int du
\frac{u^{7/4}}{\sqrt{h}}\sqrt{A(u)}\sqrt{\left(\frac{R}{u}\right)^{3/2}f^{-1}+\left(\frac{u}{R}
  \right)^{3/2}f\left(\frac{\pt x_4}{\pt u}\right)^2}, \label{actionB1} \eea
where the $A(u)$ is defined as \be
A(u)\equiv\frac{u^3}{R^3+b^2u^3}+\frac{B^2}{1+2bB}.\ee It is clear that the
factor $A(u)$ can't be canceled each other with the $h(u)$ in the action
(\ref{actionB1}) through introducing this magnetic field. If the magnetic field
vanishes $B=0$, then this effective action will reduce to the action
(\ref{action1}).

\begin{figure}[ht]
 \centering
 \includegraphics[width=0.55\textwidth]{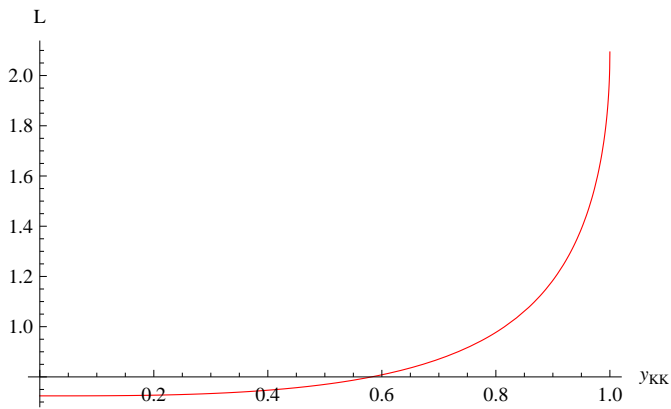}
 \caption{The asymptotic distance $L$ between the D8 and ${\rm\overline{D8}}$
   varies with $y_{KK}$ at $b,~ B=0$. }
\label{fig1}
 \end{figure}

\begin{figure}[ht]
 \centering
\begin{tabular}{cc}
 \includegraphics[width=0.45 \textwidth]{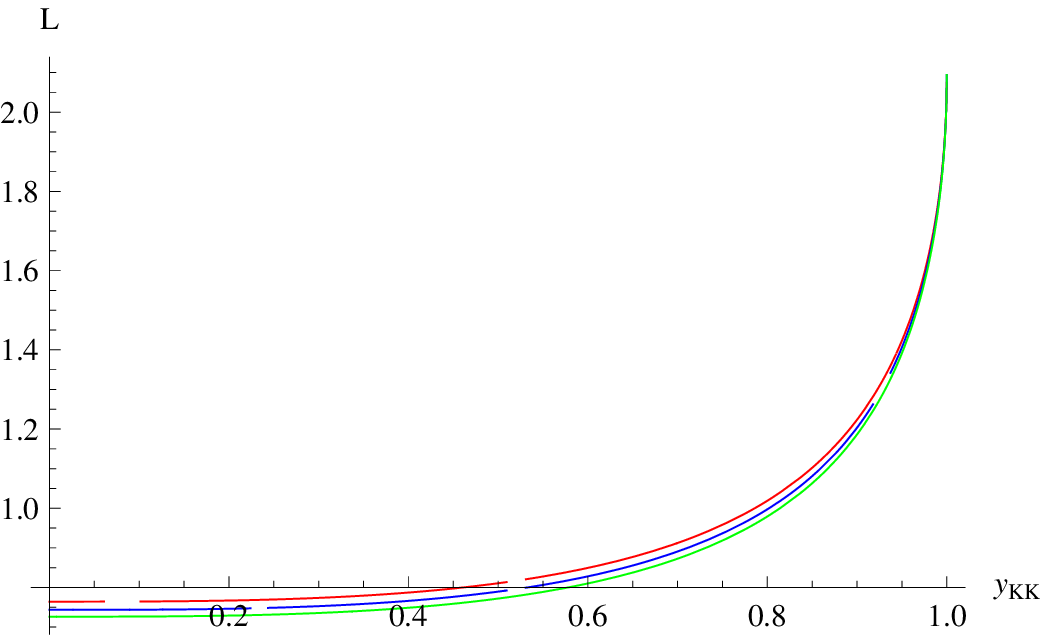}&
\includegraphics[width=0.45\textwidth]{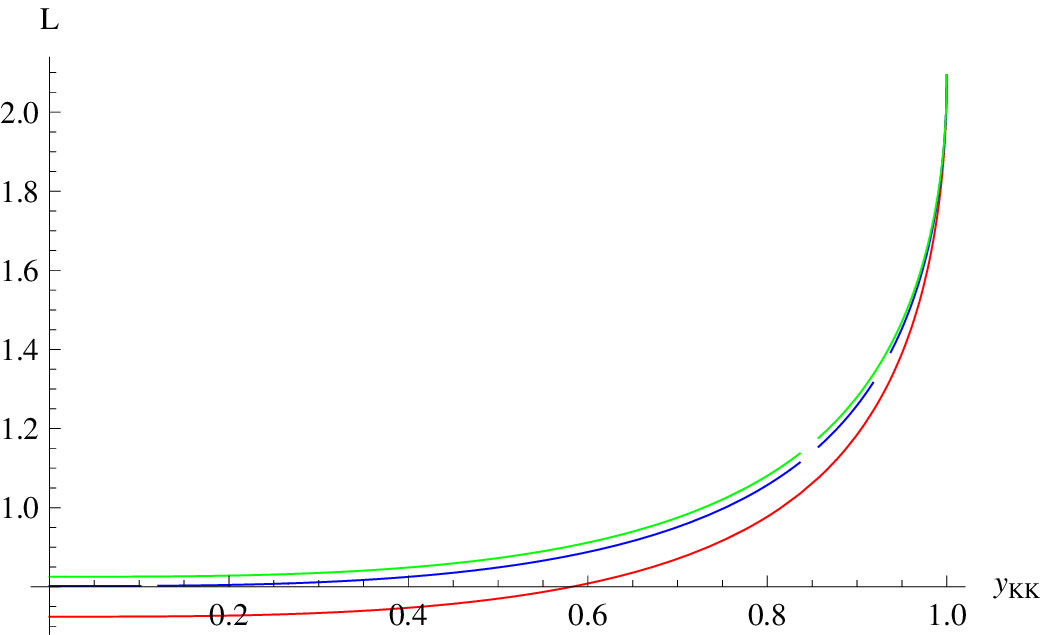}\\
(a)  & (b) 
 \end{tabular}
\caption{The asymptotic distance $L$: (a) $b=1,~2,~9$ (red, blue, green) or (from top to bottom) and $B=1$;
(b) $B=0,~3,~9$ (red, blue, green) or (from bottom to top) and $b=1$.} \label{fig2}
\end{figure}

\begin{figure}[ht]
 \centering
\begin{tabular}{cc}
\includegraphics[width=0.45\textwidth]{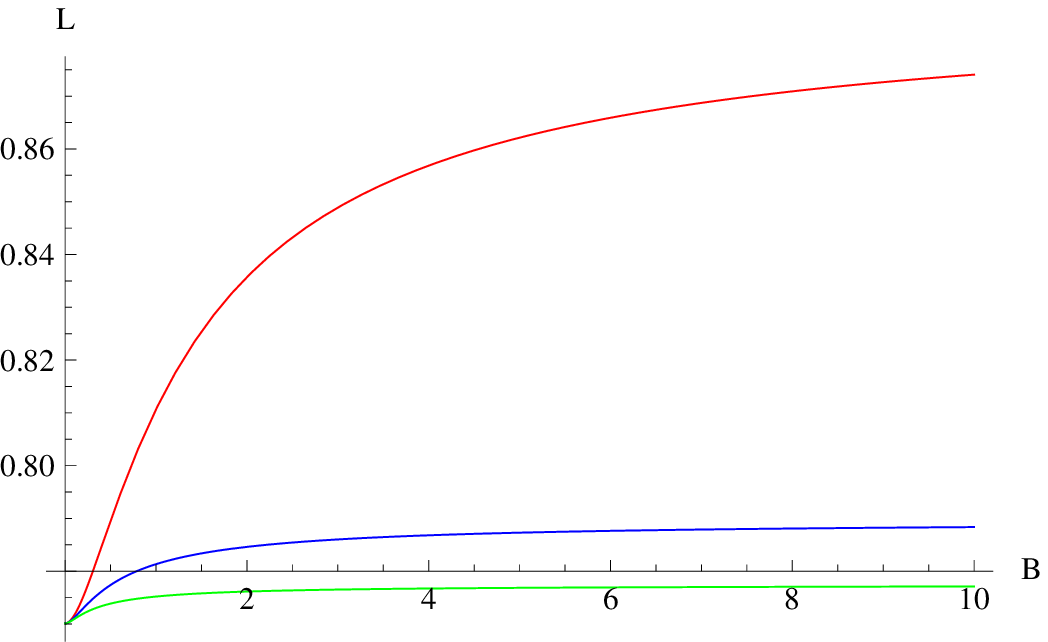}&
 \includegraphics[width=0.45\textwidth]{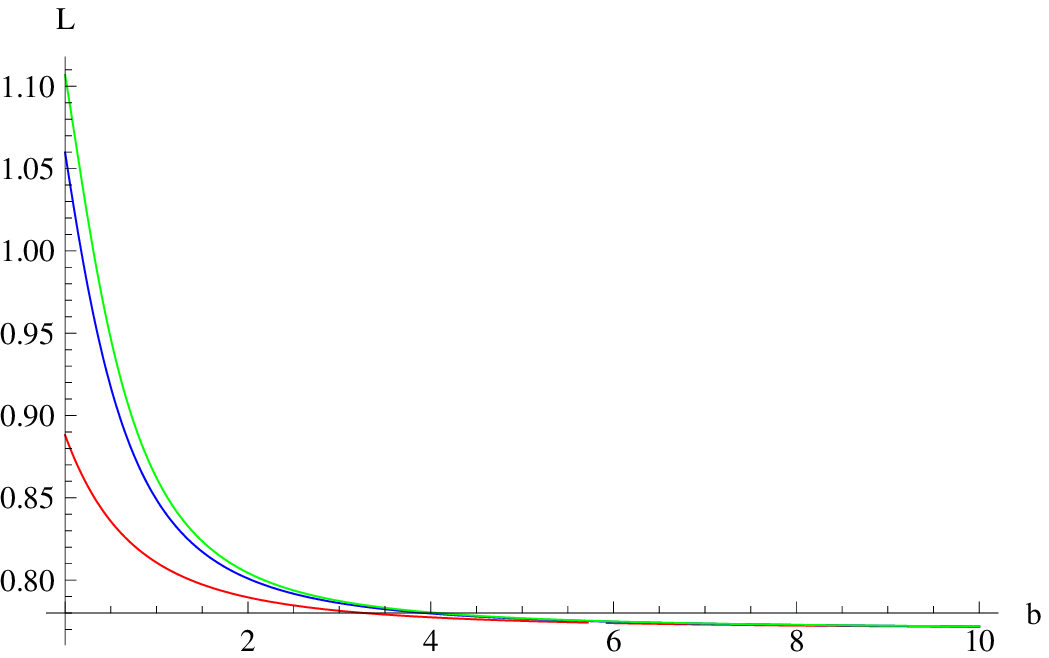}\\
 (a) & (b)
 \end{tabular}
\caption{The asymptotic distance $L$: (a) $b=1,~3,~5$ (red, blue, green) or
  (from top to bottom) 
and $y_{KK}=0.5$;
(b) $B=1,~3,~5$ (red, blue, green)  or (from bottom to top) and $y_{KK}=1$. } \label{fig3}
\end{figure}

Now the equation of motion from the action (\ref{actionB1}) is \be
\frac{\pt}{\pt
  {x_4}}\left[\frac{u^{5/2}}{\sqrt{h}}\frac{f\sqrt{A(u)}}{\sqrt{f+f^{-1}
      \left(\frac{R}{u}\right)^{3}u'^2}}\right]=0.\ee Choose a boundary
condition $u'=0$ at $u=u_0$ and perform an integration\footnote{Here the $'$
  denotes $\pt_{x_4}$.}, we get a first derivative equation\be
\frac{u^{5/2}}{\sqrt{h}}\frac{f\sqrt{A(u)}}{\sqrt{f+f^{-1}\left(\frac{u}{R}\right)^{3}u'^2}}=
\frac{u_0^{5/2}\sqrt{f(u_0)A(u_0)}}{\sqrt{h(u_0)}}.\label{eom1} \ee Then the
$u'$ equation is \bea && u'=\sqrt{P(u)},\cr &&~~ P(u)\equiv
\left(\frac{u}{R}\right)^{3}f^2\left[\left(\frac{u}{u_0}\right)^5
  \frac{h(u_0)A(u)f(u)}{h(u)A(u_0)f(u_0)}-1\right]. \eea Defining $y\equiv
\frac{u}{u_0}$, $y_{KK}\equiv \frac{u_{KK}}{u_0}$ and $z=y^{-3}$, we can rewrite
the above equation $P(u)$ as \bea
P&=&\frac{u_0^3}{R^3}y^3f(y)^2\left[y^5\frac{h(1)A(y)f(y)}{h(y)A(1)f(1)}-1\right]\cr
&=&\frac{u_0^3}{R^3}z^{-1}f(z)^2\left[z^{-5/3}\frac{h(1)A(z)f(z)}{h(z)A(1)f(1)}-1\right].\eea
So the shape of this connected D8-${\rm\overline D8}$ brane solution is \be
x_4(u)=\int_{u_0}^u \frac{du}{\sqrt{P(u)}}.\ee The corresponding chiral
symmetry in the effective field theory is broken because of this connected
solution. Then the asymptotic distance
between the D8 and ${\rm \overline{D8}}$ reads \be L=2\int_{u_0}^\infty
\frac{du}{\sqrt{P(u)}}=\frac{2u_0}{3}\int_0^1\frac{dz}{z^{4/3}\sqrt{P(z)}}.\ee
We plot some figures \ref{fig1}-\ref{fig3} to show this asymptotic distance
how to depend on the variable $b$, $B$ and $y_{KK}$ in this low temperature
phase\footnote{In this section, we alway choose
  $R=u_0=1$ in doing the numerical calculations.}.
These figures shows the asymptotic distance $L$
between the D8 and ${\rm \overline {D8}}$ brane is became larger with increasing the
value $y_{KK}$ at arbitrary value $b$ and $B$. For a fixed $B$ and increasing
$b$, then $L$ is decreased. However, for a fixed $b$ and increasing $B$, the
distance $L$ is also increased. So the magnetic field $B$ and non-commutative
parameter $b$ have a converse contribution to the asymptotic distance $L$. 

\begin{figure}[ht]
 \centering
\begin{tabular}{cc}
\includegraphics[width=0.45\textwidth]{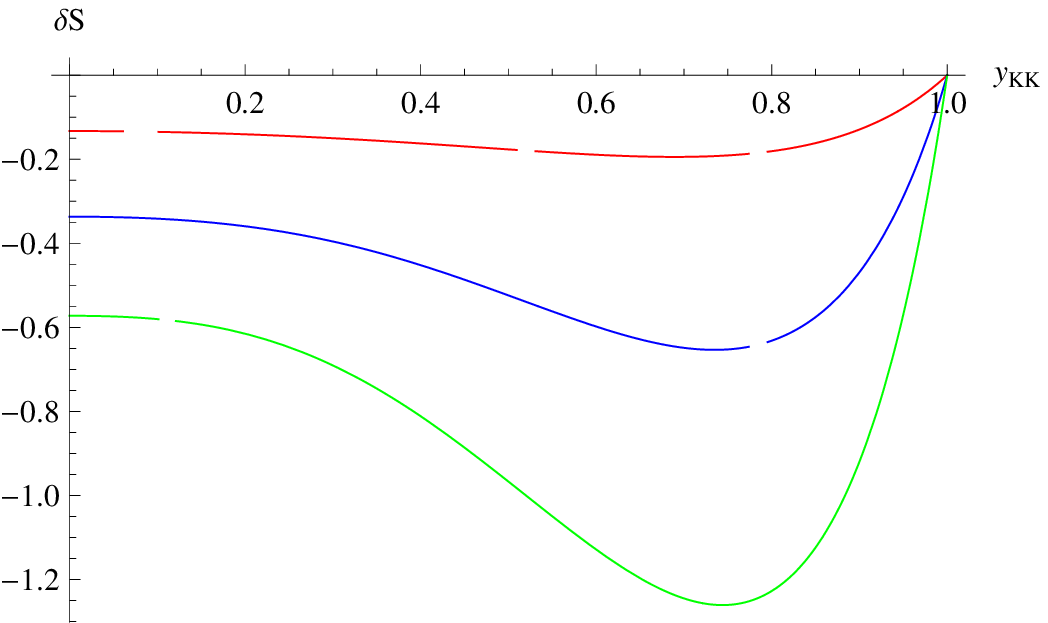}&
 \includegraphics[width=0.45\textwidth]{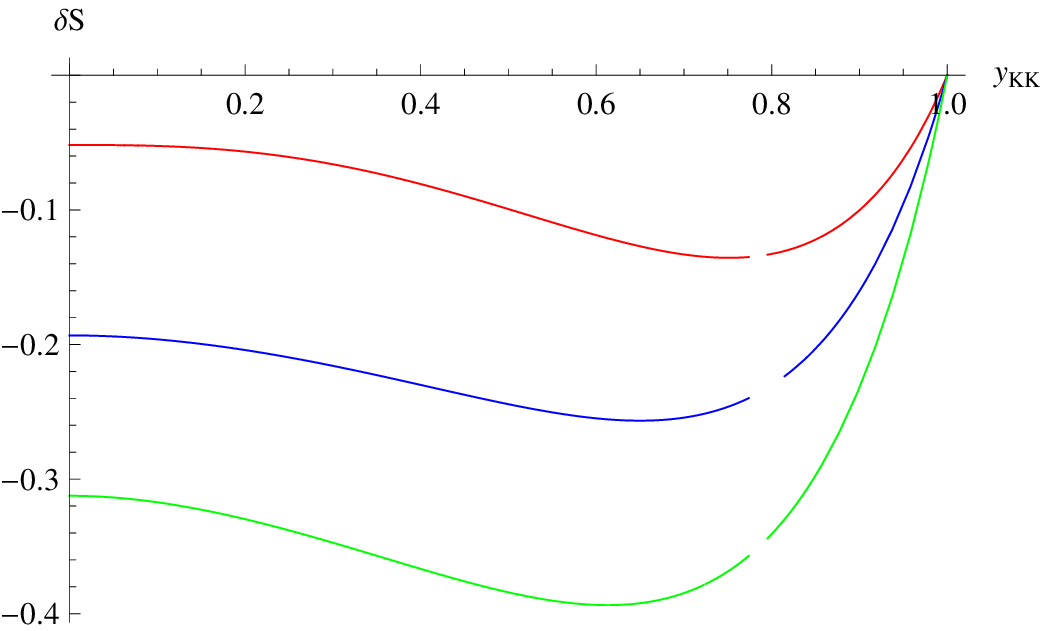}\\
  (a)   & (b) 
 \end{tabular}
\caption{The energy difference $\delta S$: (a) $b=1,~3,~5$ (red, blue, green) or
  (from top to bottom) and $B=1$;
(b) $B=0,~2,~5$ (red, blue, green) or (from top to bottom) and $b=1$. } \label{fig4}
\end{figure}

\begin{figure}[ht]
 \centering
\begin{tabular}{cc}
\includegraphics[width=0.45\textwidth]{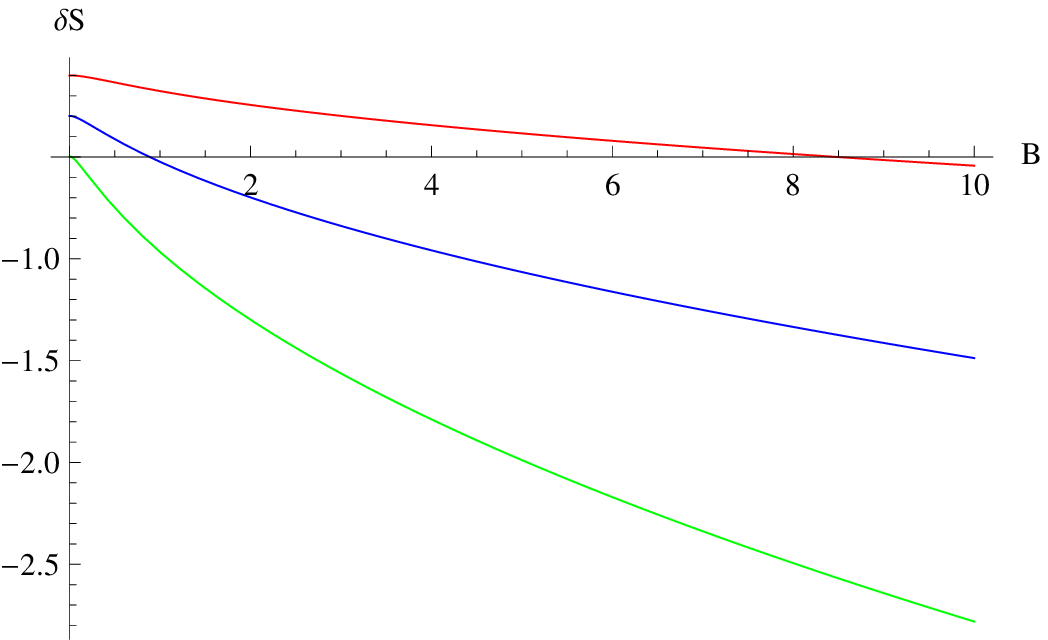}&
 \includegraphics[width=0.45\textwidth]{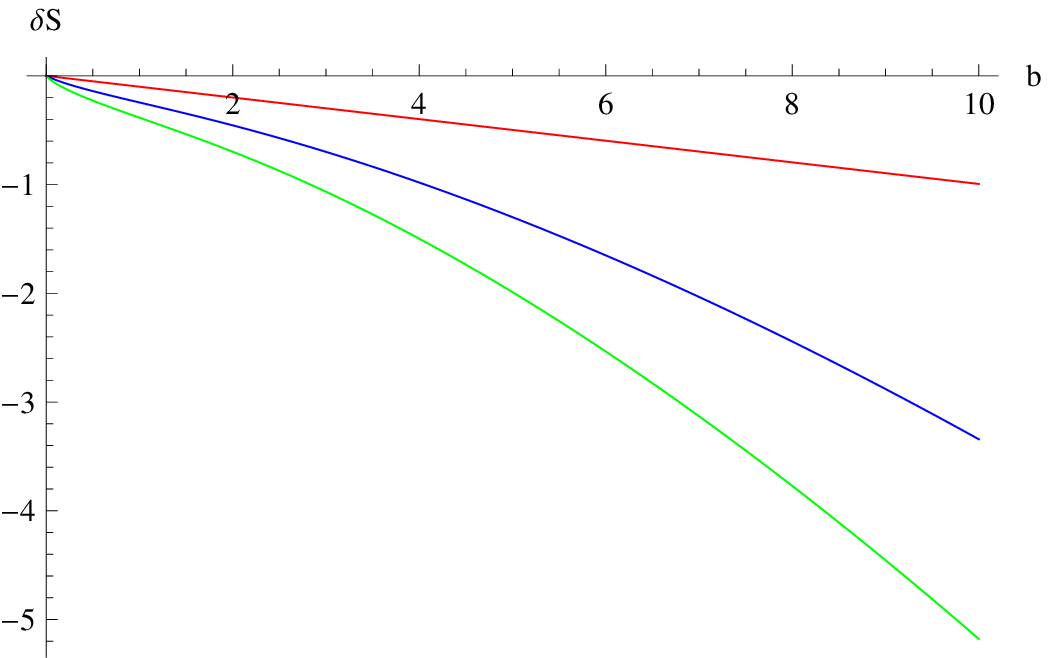}\\
  (a)  & (b) 
 \end{tabular}
\caption{The energy difference $\delta S$: (a) $b=1,~3,~5$ (red, blue, green) or
  (from top to bottom) and $y_{KK}=0.5$;
(b) $B=0,~2,~5$ (red, blue, green) or (from top to bottom) and $y_{KK}=0.5$. } \label{fig5}
\end{figure}

Substituting the equation of motion (\ref{eom1}) into the effective action
(\ref{actionB1}), we obtain the on-shell energy of this connected D8-${\rm
  \overline{D8}}$ configuration \be S_{connected}\sim \int_{u_0}^\infty du
\frac{u^{7/4}\sqrt{A(u)}}{\sqrt{h}}\sqrt{\left(\frac{R}{u}\right)^{3/2}f^{-1}+
  f\left(\frac{u}{R}\right)^{3/2}\frac{1}{P(u)}}. \ee And the energy of the
separated D8-${\rm\overline D8}$ brane solution $\pt_{u}x_4=0$ is \be
S_{separated}\sim \int_{u_{KK}}^\infty du
\frac{R^{3/4}u\sqrt{A(u)}}{\sqrt{h(u)f(u)}}. \ee Then the energy difference of
these two solutions is \bea \delta S &=& S_{connected}-S_{separated}\cr &\sim&
\int_{u_0}^\infty du
\frac{u\sqrt{A(u)}}{\sqrt{h(u)}}\left(\sqrt{f^{-1}+f\left(\frac{u}{R}\right)^{3}\frac{1}{P(u)}}
  -f^{-1/2}\right)\cr && \quad \quad~~~~ -\int_{u_{KK}}^{u_0}du
\frac{u\sqrt{A(u)}}{\sqrt{h(u)f(u)}}\cr &\sim&\frac{1}{3}\int_0^1 dz
\frac{\sqrt{A(z)}}{z^{5/3}\sqrt{h}}\left(\sqrt{f(z)^{-1}+\left(\frac{u_0}{R}\right)^{3}
\frac{f(z)}{P(z)}z^{-1}} 
  -f(z)^{-1/2}\right)\cr && \quad \quad~~~~ -\int_{y_{KK}}^1dy
\frac{y\sqrt{A(y)}}{\sqrt{h(y)f(y)}}.\eea In the figures \ref{fig4}
and \ref{fig5}, it shows this energy difference varying with the parameter $B$,
$b$ and $y_{KK}$.
We find the contribution of the parameters $B$ and $b$ are similar to the energy
difference $\delta S$. And this energy difference is always negative at
arbitrary values of the parameters $b$, $B$ and $y_{KK}$. So the connected
D8-${\rm\overline{D8}}$ brane configuration is always dominated, and the chiral
symmetry of the effective non-commutative field theory is always broken in the
low temperature.

\subsubsection{ High temperature}
If the magnetic field (\ref{magnetic}) is included, the effective D8 brane
action in the black hole background (\ref{metric6}) is 
\bea S &\sim& \int du
\frac{u^{7/4}\sqrt{A}}{\sqrt{h}}\sqrt{\left(\frac{R}{u}\right)^{3/2}+H(u)\left(\frac{u}{R}\right)^{3/2}
  \left(\frac{\pt x_4}{\pt u}\right)^2}.\label{actionB2} 
\eea 
Now the effective
action contains the parameters $B$ and $b$ through the factor $A(u)$ and
$h(u)$. Since the action (\ref{actionB2}) doesn't explicitly depend on the
coordinate $x_4$, the Hamiltonian relative to the variable $x_4$ is
conserved. So the equation of motion is 
\be \frac{\pt}{\pt
  x_4}\left[\frac{u^{5/2}H\sqrt{A(u)}}{\sqrt{h(u)\left(H+\left(\frac{R}{u}\right)^{3}u'^2\right)}}
\right] =0.\ee 
As the same way used at low temperature, we choose a boundary condition
as $u'=0$ at $u=u_0$, it corresponds to the connected D8-${\rm\overline{D8}}$
brane solution. Then we obtain a first derivative equation \be
\frac{u^{5/2}H\sqrt{A}}{\sqrt{h\left(H+\left(\frac{R}{u}\right)^{3}u'^2\right)}}=\frac{u_0^{5/2}
  \sqrt{H(u_0)A(u_0)}}{\sqrt{h(u_0)}}. \ee

\begin{figure}[ht]
 \centering
 \includegraphics[width=0.55\textwidth]{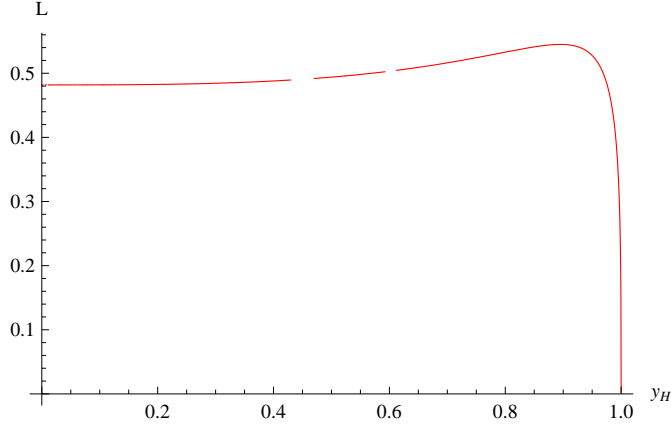}
 \caption{ The asymptotic distance $L$ is shown at $b,~B=0$. }
 \label{fig6}
\end{figure}

\begin{figure}[ht]
 \centering
\begin{tabular}{cc}
 \includegraphics[width=0.45\textwidth]{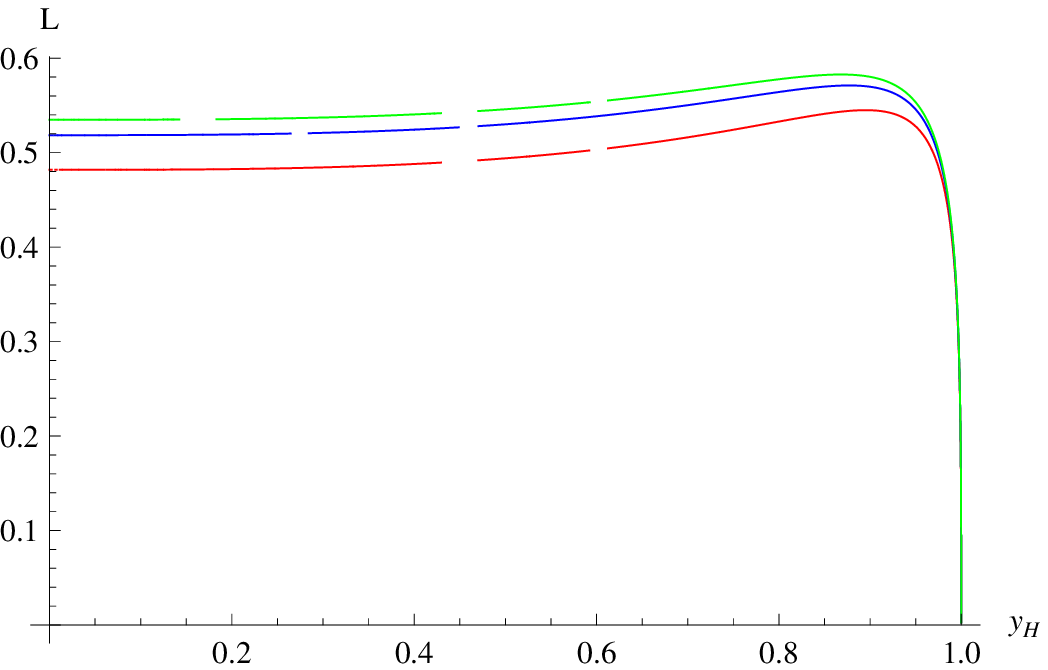}&
\includegraphics[width=0.45\textwidth]{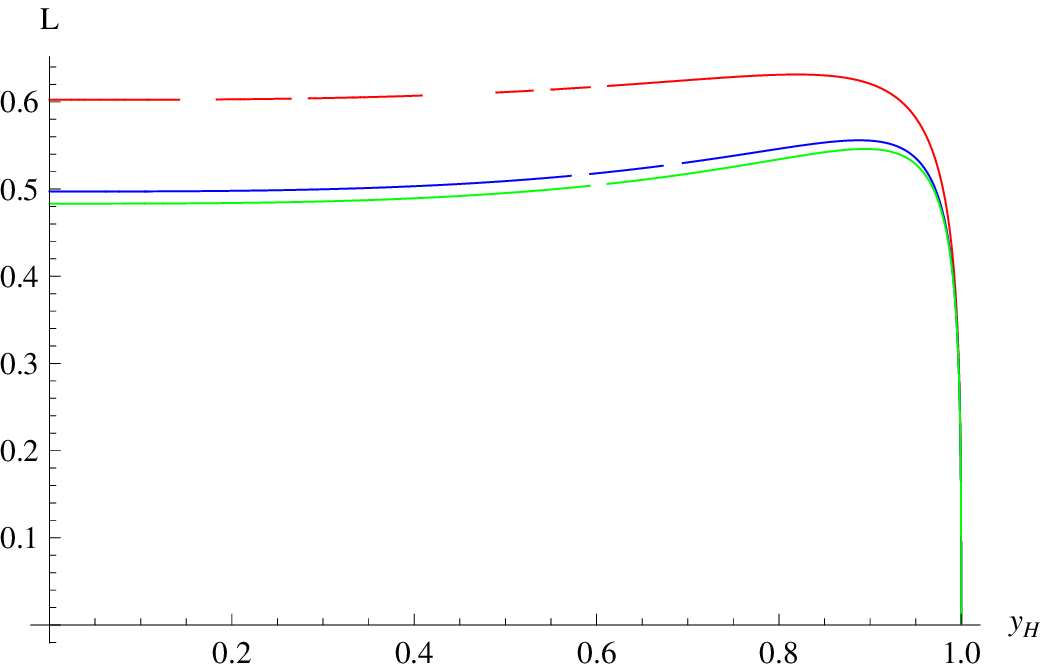}\\
   (a)   & (b) 
 \end{tabular}
\caption{The asymptotic distance $L$: (a) $B=0,~2,~6$ (red, blue, green) or (from bottom to top) and
  $b=1$; (b) $b=0,~2,~8$ (red, blue, green) (from top to bottom) and $B=2$.
 } \label{fig7}
\end{figure}

\begin{figure}[ht]
 \centering
\begin{tabular}{cc}
\includegraphics[width=0.45\textwidth]{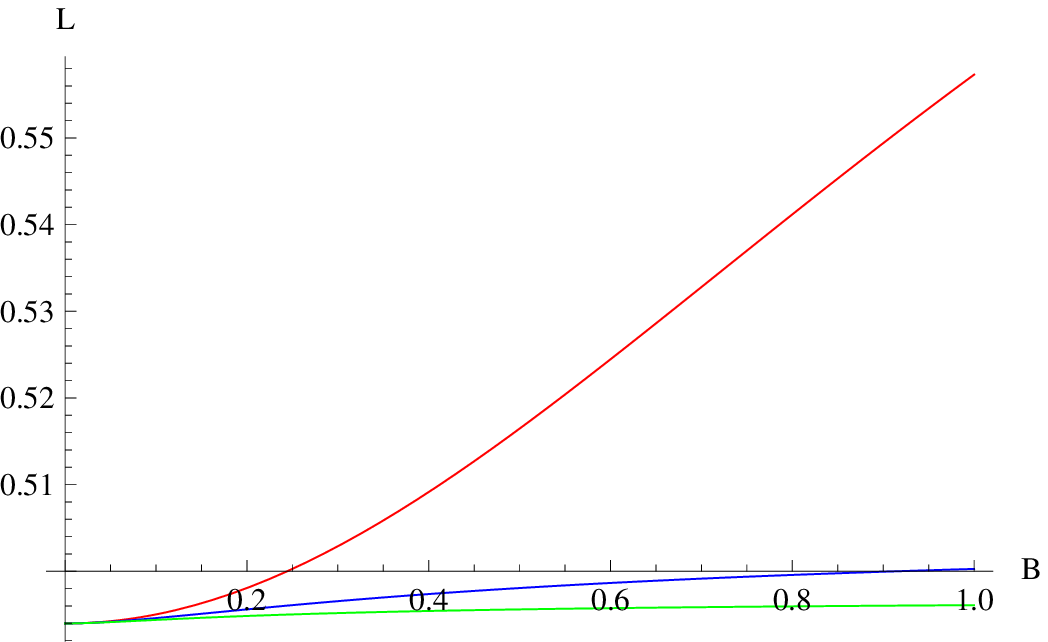}&
\includegraphics[width=0.45\textwidth]{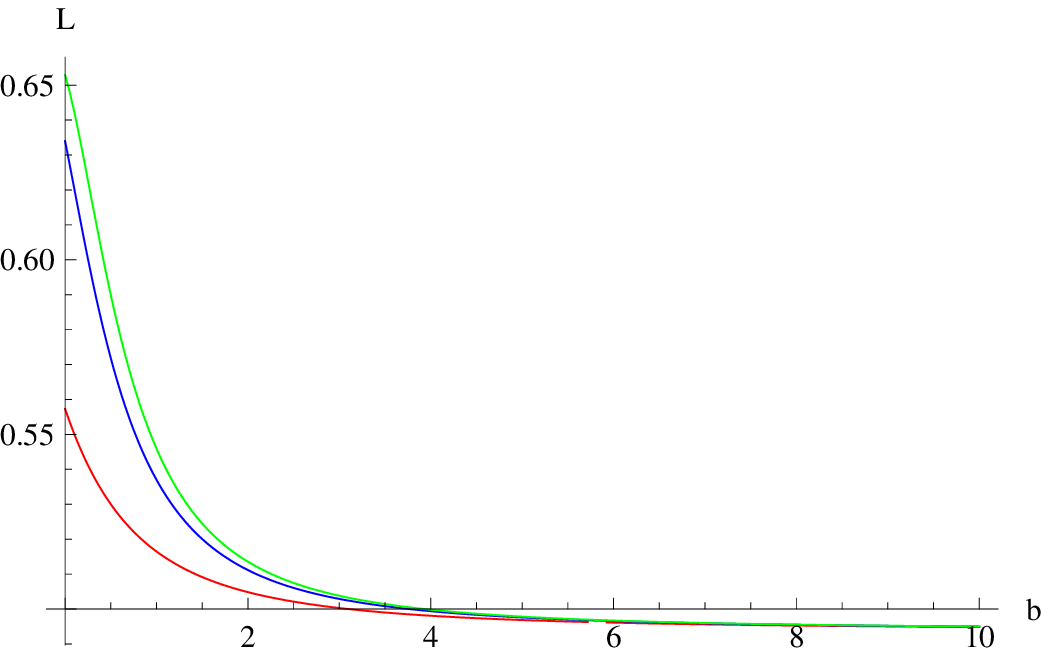} \\
   (a)   & (b) 
 \end{tabular}
\caption{The asymptotic distance $L$: (a) $b=0,~3,~6$ (red, blue, green) or (from top to bottom)  and
  $y_H=0.5$; (b) $B=1,~3,~6$ (red, blue, green) or (from bottom to top) and $y_H=0.5$.} 
\label{fig8}
\end{figure}

In the new variables $y\equiv \frac{u}{u_0}$ and $z\equiv y^{-3}$, the equation
can be written as \bea u' &=& \sqrt{Q(u)},\cr
Q(u)&\equiv&\left(\frac{u}{R}\right)^{3}\left(\frac{u^5H(u)^2A(u)h(u_0)}{h(u)A(u_0)H(u_0)u_0^5}-H(u)
\right)\cr
&=&\frac{u_0^3}{R^3}z^{-1}\left(\frac{H(z)^2A(z)h(1)}{z^{5/3}h(z)A(1)H(1)}-H(z)
\right).\eea So the asymptotic distance between the D8 and ${\rm \overline{D8}}$
defines \be L=2\int_{u_0}^\infty
\frac{du}{\sqrt{Q(u)}}=\frac{2u_0}{3}\int_0^1\frac{dz}{z^{4/3}\sqrt{Q(z)}}. \ee
Through some numerical calculations, we plot figures \ref{fig6}, \ref{fig7}
and \ref{fig8} about this asymptotic distance $L$ varying with the parameter $b$,
$B$ and $y_{H}$.
The asymptotic distance $L$ of the D8 and ${\rm \overline D8}$ brane become
larger with increasing the value $B$ and a fixed value $b$. However, for a fixed
$B$, then $L$ is decreased with increasing the value $b$. As at low temperature,
the magnetic parameter $B$ and non-commutative parameter $b$ have a converse
contribution to the asymptotic distance $L$ between D8 and ${\rm\overline{D8}}$.
 
For this connected D8-${\rm\overline{D8}}$ brane solution, after inserting the
solution into the effective action (\ref{actionB2}), we obtain its on-shell energy
\be S_{connected}\sim \int_{u_0}^\infty
\frac{u^{7/4}A(u)}{\sqrt{h(u)}}\sqrt{\left(\frac{R}{u}\right)^{3/2}+\left(\frac{u}{R}\right)^{3/2}
  \frac{H(u)}{Q(u)}}.\ee Similar as the low temperature case, there also exists a
separated D8-${\rm\overline{D8}}$ solution, and its energy is \be
S_{separated}\sim \int_{u_H}^\infty du \frac{u^{7/4}A(u)}{\sqrt{h(u)}}
\left(\frac{R}{u}\right)^{3/4}.\ee

\begin{figure}[ht]
 \centering
 \includegraphics[width=0.55\textwidth]{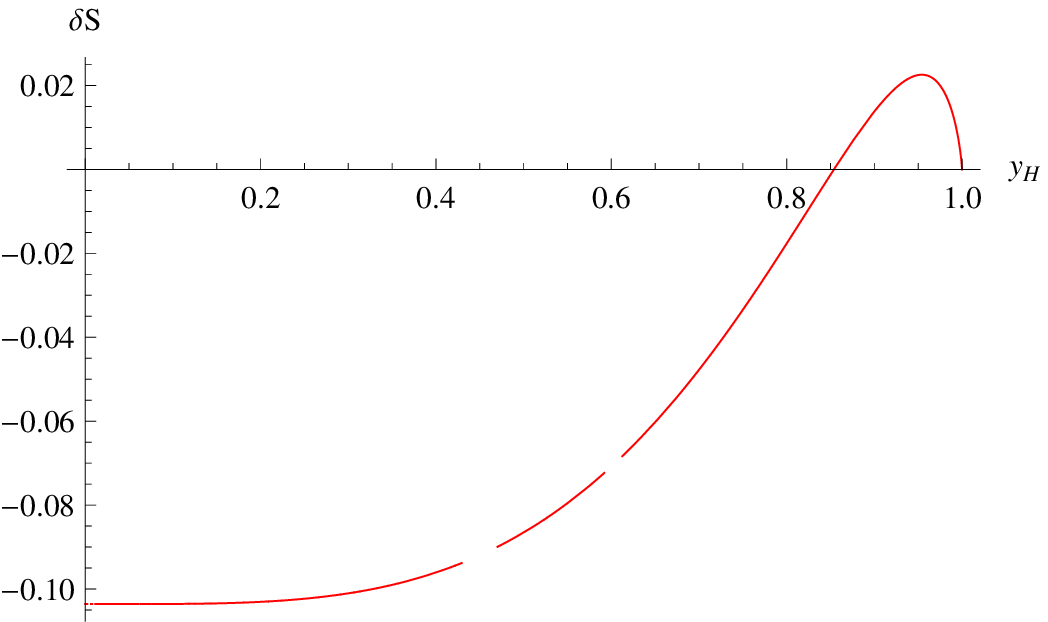}
 \caption{ This figure is the energy difference $\delta S$ at $B=0$ and $b=0$.  }
 \label{fig9}
 \centering
\begin{tabular}{cc}
\includegraphics[width=0.46\textwidth]{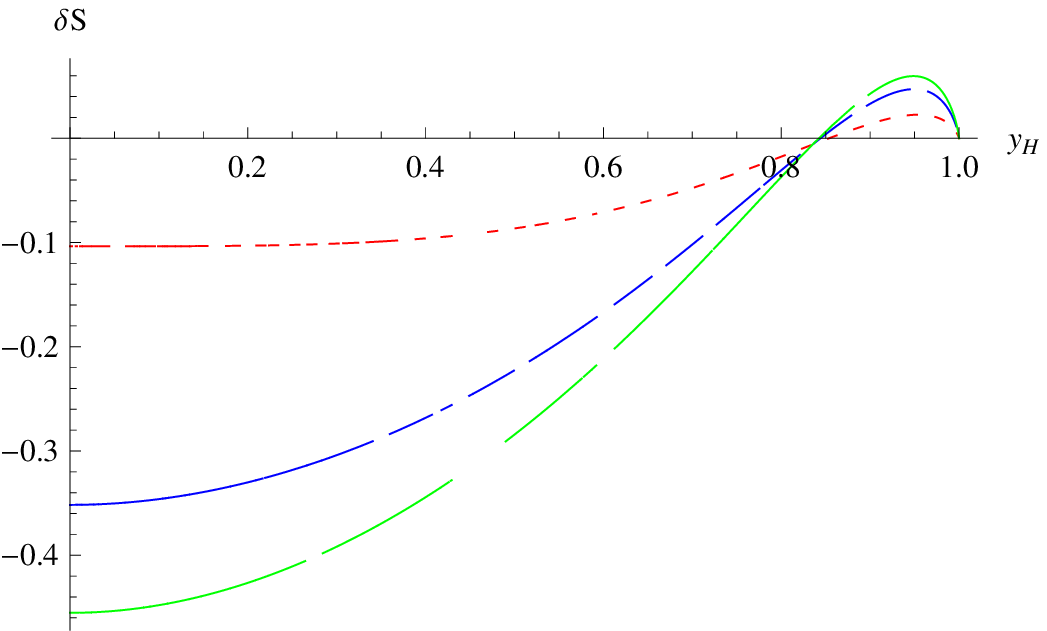}&
\includegraphics[width=0.46\textwidth]{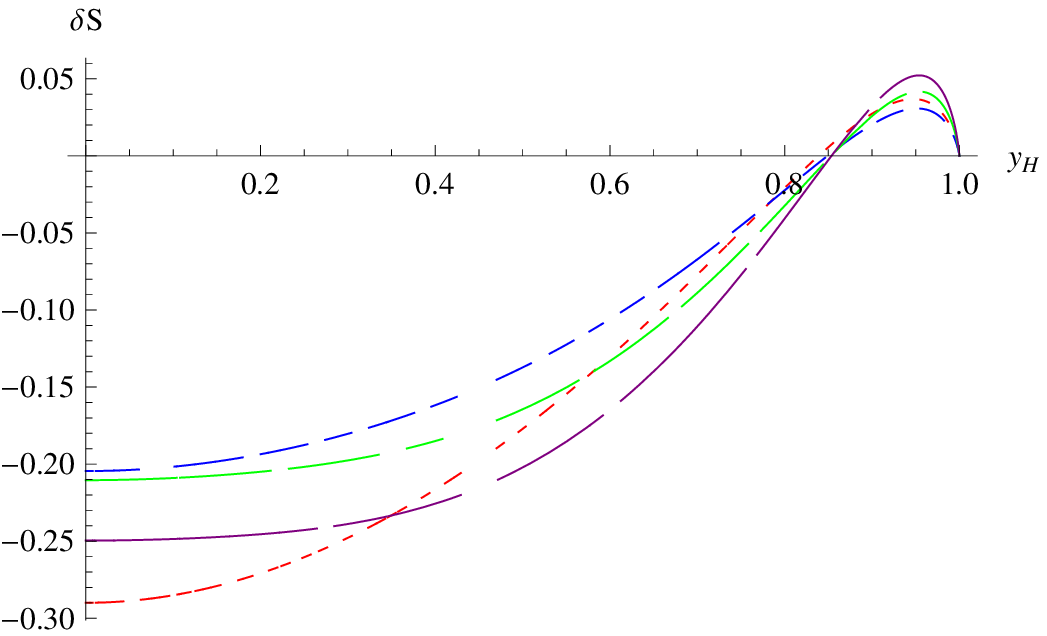} \\
   (a)   & (b) 
 \end{tabular}
\caption{It is the energy difference $\delta S$: (a) at $B=0,~2,~5$ (red, blue,
  and green) or (the length of the dashed line segment is increased) and $b=1$; 
 (b) at $b=0,~1,~5,~9$ (red, blue, green and purple) or (the length of the dashed 
 line segment becomes larger) and $B=1$. } 
\label{fig10}
\end{figure}

So the difference of the energy is \bea \delta
S&=&S_{connected}-S_{separated}\cr &\sim & \int_{u_0}^\infty du
\frac{u\sqrt{A(u)}}{\sqrt{h(u)}}\left(\sqrt{1+
    \left(\frac{u}{R}\right)^{3}\frac{H(u)}{Q(u)}}-1\right)\cr && \quad \quad
~~-\int_{u_H}^{u_0}du \frac{u\sqrt{A(u)}}{\sqrt{h(u)}}\cr &\sim
&\frac{1}{3}\int_0^1 dz
\frac{\sqrt{A(z)}}{z^{5/3}\sqrt{h(z)}}\left(\sqrt{1+\frac{u_0^3}{R^3}\frac{H(z)}{Q(z)}z^{-1}}-1\right)
\cr && \quad \quad ~~-\int_{y_H}^1 dy \frac{y\sqrt{A(y)}}{\sqrt{h(y)}}.\eea We explicitly
show this
energy difference $\delta S$ in the figures
\ref{fig9} and \ref{fig10}.
 
It is clear that the energy difference between the connected and separated
solutions has critical point ${y_{H}}^c$, i.e. critical temperature $T_\xi$. If
$B=b=0$, this holographic model becomes the Sakai-Sugimoto model. This chiral phase
transition is studied in \cite{Aharony:2006da}. At $b=0$
and a finite $B$, the chiral phase transition is also investigated in
\cite{Johnson:2008vna}. The results got here at $B=b=0$ or $b=0$ is all same as
in \cite{Aharony:2006da} and \cite{Johnson:2008vna}. At
a fixed value $b$, the critical point value $y_H$ becomes smaller with
increasing the value $B$ in figure \ref{fig10}(a). However, it become larger with
increasing the value $b$ for a fixed $B$ in figure \ref{fig10}(b). The influences of
the parameters $B$ and $b$ is very little. Just like the effects
on the asymptotic distance, these two parameters also have converse
contributions to the critical point of phase transition. 

Below these critical points, the energy difference $\delta S$ is negative, otherwise,
it become positive. It means there exists two phases at high temperature. The
phase transition is first-order between each other. Under the value ${y_{H}}^c$,
the connected solution is dominated, then the chiral symmetry in the
four-dimensional field theory will be broken $U(N_f)\times U(N_f)\rightarrow
U(N_f)_{\rm diag}$. Otherwise, the dominated solution is the separated
D8-${\rm\overline{D8}}$ configuration, and the chiral symmetry will be restored.

\section{General brane configuration}
In this section, we generalize to consider some other brane configurations. Here
we mainly consider the color brane still is the D4-branes, so the corresponding
gravity backgrounds are the equations (\ref{metric5}) and (\ref{metric6}).

Firstly, let us consider the following brane constructions  
\be
\begin{array}{ccccccccccccc}
&0 &1 &2 &3 &4 &5 &6 &7 &8 &9\\
N_c~{\rm D4}: &{\rm x} &{\rm x} &{\rm x} &{\rm x} &{\rm x} &{} &{} &{} &{} &{}\\
N_f~{\rm D6,\overline{D6}}: &{\rm x} &{\rm x} &{\rm x} &{\rm x} &{} &{\rm x}
&{\rm x} &{\rm x} &{ }  &{ }\\
N_f~{\rm D4,\overline{D4}}: &{\rm x} &{\rm x} &{\rm x} &{\rm x} &{} &{\rm x}
&{} &{} &{}  &{}
\end{array}\label{configuration2}
\ee The effective field theory on the intersecting region is some
four-dimensional non-commutative theory. Follow the same method in the previous
subsection, we can analyze the dynamics of these non-commutative field theory at
strong coupled regime by supergravity approximation. The effective action of the
flavor brane in the low temperature background (\ref{metric5}) and black hole
background (\ref{metric6}) are as follows \bea &&S_{\rm low}\sim \int du
u^{n}\sqrt{\left(\frac{R}{u}\right)^{3/2}f^{-1}+\left(\frac{u}{R}\right)^{3/2}\left(\frac{\pt
      x_4}{\pt u}\right)^2},\\
&& S_{\rm High}\sim \int du
u^{n}\sqrt{\left(\frac{R}{u}\right)^{3/2}+H(u)\left(\frac{u}{R}\right)^{3/2}\left(\frac{\pt
      x_4}{\pt u}\right)^2}.\eea For the flavor D6-brane, the parameter $n$ is $11/4$,
while $n=9/4$ for the D4-brane. The non-commutative parameter $b$ is canceled in
the effective action. So the classical dynamics of flavor brane is same as the
commutative cases. At low temperature, the chiral symmetry is always
broken. However, there exists a chiral phase transition in the high temperature
phase. Below a critical temperature, the chiral symmetry is broken, it
corresponds to the connected D8-${\rm\overline{D8}}$ brane solution. Otherwise,
the solution is a separated brane configuration, and this symmetry is restored.

Also some three-dimensional non-commutative effective field theories can be
constructed through the following brane configurations 
\be
\begin{array}{ccccccccccccc}
 &0 &1 &2 &3 &4 &5 &6 &7 &8 &9\\
N_c~{\rm D4}: &{\rm x} &{\rm x} &{\rm x} &{\rm x} &{\rm x} &{} &{} &{} &{} &{}\\
N_f~{\rm D6,\overline{D6}}: &{\rm x} &{\rm x} &{\rm x} &{ } &{ } &{\rm x}
&{\rm x} &{\rm x} &{\rm x }  &{ }\\
N_f~{\rm D4,\overline{D4}}: &{\rm x} &{\rm x} &{\rm x} &{ } &{} &{\rm x}
&{\rm x} &{} &{}  &{}
\end{array}\label{configuration3}
\ee Now the flavor D6 and D4-brane effective actions at low and high temperature
background read \bea &&S_{\rm low}\sim \int du u^m
\sqrt{\left(\frac{R}{u}\right)^{3/2}f^{-1}+\left(\frac{u}{R}\right)^{3/2}\left(\frac{\pt
      x_4}{\pt u}\right)^2}, \\
&&S_{\rm high} \sim \int du
u^{m}\sqrt{\left(\frac{R}{u}\right)^{3/2}+H(u)\left(\frac{u}{R}\right)^{3/2}\left(\frac{\pt
      x_4}{\pt u}\right)^2}, \eea where the parameter $m$ is equal to $9/4$ for
the D6-brane, and is $7/4$ for the D4-brane. Again these effective action are
same as the commutative cases.

As the same discussions in the above subsections, one can turn on a magnetic field
along the directions $x_1$ and $x_2$ on the worldvolume of the flavor
branes. Then in the flavor D-brane effective action there will exist some
contributions of the $B$ and $b$ through $A(u)$ and $h(u)$. Here we list these
effective actions: (a) for brane configurations (\ref{configuration2}), the
corresponding effective actions are \bea
&&S_{\rm low_{bB}}\sim \int du
\frac{u^{n-3/2}}{\sqrt{h(u)}}\sqrt{A(u)}\sqrt{\left(\frac{R}{u}\right)^{3/2}f^{-1}+\left(\frac{u}{R}
  \right)^{3/2}\left(\frac{\pt
      x_4}{\pt u}\right)^2},\\
&& S_{\rm High_{bB}}\sim \int du
\frac{u^{n-3/2}}{\sqrt{h(u)}}\sqrt{A(u)}\sqrt{\left(\frac{R}{u}\right)^{3/2}+H(u)\left(\frac{u}{R}
  \right)^{3/2}\left(\frac{\pt x_4}{\pt u}\right)^2}; \eea (b) for brane
configurations (\ref{configuration3}), the actions read
 \bea &&S_{\rm low_{bB}}\sim \int du
\frac{u^{m-3/2}}{\sqrt{h(u)}}\sqrt{A(u)}
\sqrt{\left(\frac{R}{u}\right)^{3/2}f^{-1}+\left(\frac{u}{R}\right)^{3/2}\left(\frac{\pt
      x_4}{\pt u}\right)^2}, \\
&&S_{\rm high_{bB}} \sim \int du
\frac{u^{m-3/2}}{\sqrt{h(u)}}\sqrt{A(u)}\sqrt{\left(\frac{R}{u}\right)^{3/2}+H(u)\left(\frac{u}{R}
  \right)^{3/2}\left(\frac{\pt x_4}{\pt u}\right)^2}. \eea Then we can investigate
some effects of the NS-NS $B_{12}$ and magnetic field $B$ on the string coupled dynamics of the
non-commutative effective field theory through using gauge/gravity
correspondence. One will find some similar results as the previous
subsection. And maybe these holographic models can be used to study some
condensed matter physics, for example the quantum Hall effect \cite{Susskind:2001fb}. 

Also we can generalize to consider some other color brane background with a
NS-NS field, and construct the general Dq/Dp-${\rm\overline{Dp}}$ brane
configurations. Then one can study some influences of this NS-NS background
field on some properties of the effective theories living on the intersecting
parts of these brane configurations by the AdS/CFT correspondence.

\section{String in non-commutative background}
In this section, we consider the dynamics of a fundamental sting in the high
temperature phase. We calculate the drag force of a quark moving through the
QGP, and also study the Regge trajectory of a meson in this non-commutative QGP.

\subsection{Quark in non-commutative QGP}
The Nambu-Goto action of a fundamental string is \be
S=-\frac{1}{2\pi\alpha'}\int d\tau d\sigma \sqrt{-\det
  g_{\alpha\beta}}+\frac{1}{2\pi\alpha'}\int P[B],
~~~g_{\alpha\beta}\equiv \pt_\alpha X^\mu\pt_\beta X^\nu
G_{\mu\nu}.\ee We consider the two endpoints of a fundamental string separately
attached on the black horizon and flavor D8-brane. Following the same way in
\cite{Gubser:2006bz} and \cite{Matsuo:2006ws}, we parameterize the world-sheet
coordinates of this fundamental string as $\tau=t$ and $\sigma=u$, and assume
the endpoint (quark) on the flavor brane moving along the direction $x_2$ with
\be x_2=vt+\xi(u).\ee Since there exists a rotational symmetry in the $x_1$ and
$x_2$ plane, it is equivalent to let quark moving along the direction $x_1$.

So the induced metric on the string worldsheet is \bea &&
g_{\tau\tau}=-\left(\frac{u}{R}\right)^{3/2}(H-hv^2),
~~g_{\tau\sigma}=g_{\sigma\tau}=\left(\frac{u}{R}\right)^{3/2}hv\xi',\cr &&\quad
~~~g_{\sigma\sigma}=\left(\frac{u}{R}\right)^{3/2}h\xi'^2+\left(\frac{R}{u}\right)^{3/2}H^{-1},\eea
where the $'$ denotes $\pt_u$. Then inserting it into the string action, we get
\bea S&=&-\frac{1}{2\pi\alpha'}\int dt du \mathcal{L},\cr \mathcal{L} &\equiv&
\sqrt{1-hH^{-1}v^2+\left(\frac{u}{R}\right)^3hH\xi'^2}.\label{stringaction}\eea
The NS-NS $B_{12}$ field is a two-form along the directions $x_1$ and $x_2$. Since in
the coordinate parameterizations the coordinate $x_1$ is independent of the
string world-sheet coordinates $\tau$ and $\sigma$, this NS-NS field doesn't
have any contributions to the string action. The string world-sheet momentum of
string world-sheet is defined as \be P^r_{x_2}=\frac{\pt
  \mathcal{L}}{\pt(\pt_rx_2)}=\Pi_\xi, \ee where the canonical momentum is
$\Pi_\xi=\pt_{\xi'}L$. The string momentum is the energy associated by the
fundamental string. So the drag force is \be -f=\frac{1}{2\pi\alpha'}\Pi_\xi.\ee

\begin{figure}[ht]
 \centering
 \includegraphics[width=0.6\textwidth]{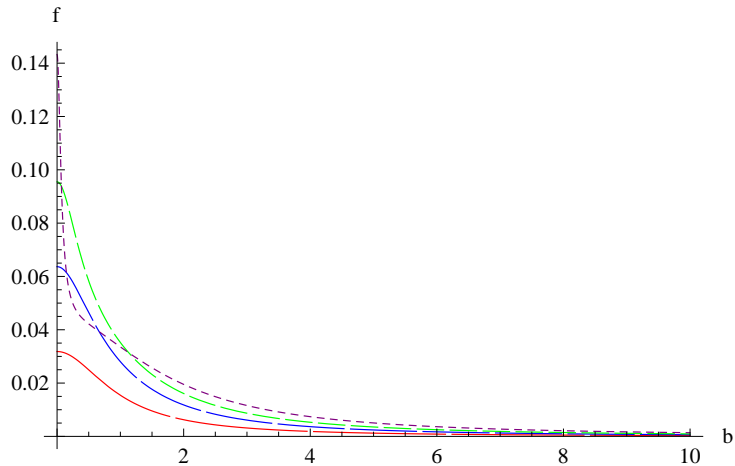}
 \caption{The drag force varies with the parameter $b$ at
   $v=0.2,~0.4,~0.6,~0.9$ (red, blue, green and purple) or 
 (the length of the dashed line segment is decreased).
 Here we already choose $R=u_H=1$ in plotting this figure. }
\label{fig11}
\end{figure}

Since the string action (\ref{stringaction}) doesn't explicitly depend on the variable
$\xi$, the canonical momentum $\Pi_\xi$ is conserved relative to the parameter
$\xi$. Then the equation of motion is derived as \be \xi'=\pm\Pi_\xi
\left(\frac{R}{u}\right)^{3/2}\sqrt{\frac{1-hH^{-1}v^2}{\left(\frac{u}{R}\right)^3h^2H^2-hH\Pi_\xi^2}}.\label{stringequation}
\ee Here we need to choose "+" equation because the fundamental string is
received a drag force. The $h(u)$ is positive, and $H(u)$ is 0 at the horizon and
is equal to 1 at infinity. So to preserve the quantity to be positive in the
square root in the equation (\ref{stringequation}), the canonical momentum need
to satisfy \be \Pi_\xi=\left(\frac{u_c}{R}\right)^{3/2}h(u_c)v, \ee where the
coordinate $u_c$ is chosen as \be
u_c^3=\frac{1}{2a^3}\left(-(1-a^3u_H^3-v^2)+\sqrt{(1-a^3u_H^3-v^2)^2+4a^3u_H^3}\right). \ee
Thus we obtain the drag force is \be
f=\frac{1}{2\pi\alpha'}\left(\frac{u_c}{R}\right)^{3/2}h(u_c)v.\ee In the figure
\ref{fig11}, it shows this drag force how to depend on the non-commutative
parameter $b$.
At a fixed velocity $v$, the drag force will be decreased with increasing the
value $b$. And at some large $b$, the drag force will vanish. Thus, the NS-NS
$B_{12}$ field can decrease the viscosity of the non-commutative QGP to a moving quark.

If the parameter $a$ is very small, then the $u_c$ can be expanded as \be
u_c^3=\frac{u_H^3}{1-v^2}\left(1-\frac{v^2u_H^3}{(1-v^2)^2}a^3\right) +
\mathcal{O}(a).\ee So at the leading order, the drag force becomes \be
f=\frac{1}{2\pi\alpha'}\left(\frac{u_H}{R}\right)^{3/2}
\frac{v}{\sqrt{1-v^2}}\left(1-\frac{v^2u_H^3}{(1-v^2)^2}a^3\right). \label{force}\ee
And if let the non-commutative parameter $a^3=\frac{b^3}{R^3}=0$, this drag
force reduces to the commutative case.

\subsection{Meson in non-commutative QGP}
Now we consider a meson moving through the non-commutative hot
QGP. Here the black hole background is the metric (\ref{metric6}). Defined
$\rho^2=x_1^2+x_2^2$, then this gravity background is written as \bea
&&ds^2=\left(\frac{u}{R}\right)^{3/2}\left[-Hdt^2+h(d\rho^2+\rho^2d\varphi^2)+dx_3^2+dx_4^2\right]
+\left(\frac{R}{u}\right)^{3/2}H^{-1}du^2,\cr
&&~~B_{\rho\varphi}=B_{12}\rho.\label{generalmetric} \eea where the $S^4$ part
is omitted.  To study a meson
moving through the QGP with a velocity $v$, we instead to choose the meson
to be rest, while boost a QGP wind along the direction $x_3$ with a constant
velocity $v$ \cite{Liu:2006nn} and \cite{Liu:2006he}. Then the final metric
takes the following form \be ds^2 = -Adt^2 + 2B dt dx_3 + C dx_3^2+ g_{xx}h\left(d\rho^2
+\rho^2d\varphi^2\right)+g_{uu}du^2,
\label{generalbackground}\ee where we have defined 
\bea && g_{xx}=\left(\frac{u}{R}\right)^{3/2},~
~H=1-K, ~ g_{uu}=\left(\frac{R}{u}\right)^{3/2}H^{-1},\cr
&& A=g_{xx}(1-K\cosh^2\eta), ~B=g_{xx}K\sinh\eta\cosh\eta, ~C=g_{xx}(1+K
\sinh^2\eta), \label{generalbackground1}\\
&& \sinh\eta = \gamma v, ~~\cosh\eta = \gamma,~~ \gamma=1/\sqrt{1-v^2}.\nn 
\eea

Now we study the open string in the background (\ref{generalbackground}) and
(\ref{generalbackground1}). Here we focus on the chiral symmetry breaking phase
at high temperature. The two endpoints of this open string attach on the  
connected D8-${\rm\overline{D8}}$ flavor brane. Then the worldsheet coordinates
of the open spinning string can be parameterized as follows \be \tau=t,
~~\sigma=\rho, ~~\varphi=\omega t,
~~x_3(\sigma),~~u(\sigma).\label{parameterization1}\ee If we set the dipole
distance of $q\bar{q}$ pair in the boundary is $L$, and the dipole is relative
to the direction $x_3$ at an angle $\theta \in [0, \pi/2]$ in the $(\rho, x_3)$
plane, then the $\sigma$ is chosen in the range
$-{{L}\over{2}}\sin\theta\leq \sigma\leq {{L}\over{2}}\sin\theta$. And the boundary
conditions for $x_3(\rho)$ is taken as $x_3(\pm {{L}\over{2}}\sin\theta)=
\pm{{L}\over{2}}\cos\theta$. For the variable $u(\rho)$, its boundary
value satisfies $u(\pm{{L}\over{2}}\sin\theta)\rightarrow\infty$. Thus, the
dynamics of open spinning string can be described by the Nambu-Goto action \be
S_{NG}=-{{1}\over{2\pi\alpha'}}\int d\tau d\sigma\sqrt{-\det
  g_{\alpha\beta}}+\frac{1}{2\pi\alpha'}\int P[B] \label{NG}\ee with the induced
metric $g_{\alpha\beta}=G_{\mu\nu}\partial_\alpha x^\mu\partial_\beta x^\nu$ on
the string worldsheet \be g_{\tau\tau}= -A+g_{xx}h\rho^2\omega^2,
~~g_{\tau\sigma}=Bx_3',~~ g_{\sigma\sigma}=g_{xx}h+ Cx_3'^2+g_{uu}u'^2.\ee Its
determinant and the pullback of the NS-NS $B_{12}$ field are \bea -g_1&\equiv&
-\det
g_{\alpha\beta}=g_{xx}(1-K\cosh^2\eta-h\rho^2\omega^2)(g_{xx}h+g_{uu}u'^2)+
\nonumber~~~~~\\ &&~~~~~~
g_{xx}^2[1-K-(1+K\sinh^2\eta)h\rho^2\omega^2]x_3'^2,\cr &&P[B]=B_{12}\rho\omega
d\sigma d\tau,
\label{determinant}\eea where the sign $'$ in the formula denotes
$\partial_\sigma$. From the formulas (\ref{NG}) and (\ref{determinant}),
the equations of motion for the variables $x_3$ and $u$ read \bea
p&=&{{g_{xx}^2[1-K-(1+K\sinh^2\eta)h\rho^2\omega^2]}\over{\sqrt{-g_1}}}x_3'
\label{equation1}\\
\partial_u(\sqrt{-g_1}-B_{12}\rho\omega)&=&\partial_\sigma
\left({{g_{xx}g_{uu}(1-K\cosh^2\eta-h\rho^2\omega^2)u'}\over{\sqrt{-g_1}}}\right)
\label{equation2}\eea where $p$ is a integral constant. Since the
Lagrangian explicitly depends on the variable $\sigma$, it means that the
Hamiltonian is not conserved with respect to the variable $\sigma$. Also the
equation (\ref{equation2}) explicitly depends on the variable $\sigma$, the
equations (\ref{equation1}) and (\ref{equation2}) are difficult to analytically
solve. If the angular velocity $\omega$ is equal to zero, then the pullback of the
NS-NS $B_{12}$ field and the determinant of the induced metric (\ref{determinant})
don't explicitly depend on this variable. In this case, there exists two
conserved charges. One is the Hamiltonian relative to the variable $\sigma$, the
other is the $p$ in the equation (\ref{equation1}). Such meson $q\bar{q}$ in QGP
has been extensively studied in \cite{Liu:2006nn}-\cite{Caceres:2006ta}. And
there are not drag force to the meson moving through the QGP by the arguments
\cite{Peeters:2006iu} and \cite{Chernicoff:2006hi}. From equation (\ref{equation2}), if
$\omega=0$, then it is explicitly symmetric under the transformation
$\sigma\leftrightarrow-\sigma$. The $u'(\sigma)$ decreases in the range
$-{{L}\over{2}}\sin\theta<\sigma<0$, and is equal to zero at the critical point
$\sigma=0$, then increases to infinity in the range
$0<\sigma<{{L}\over{2}}\sin\theta$.
 
\begin{figure}[ht]
 \centering
 \includegraphics[width=0.6\textwidth]{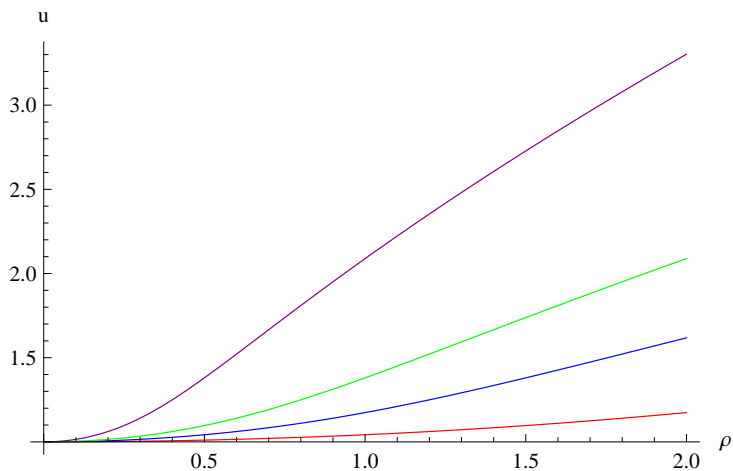}
 \caption{This figure shows the bound for the $u(\rho)$ at $\omega=0.5,~1,~1.5,~3$ (red, blue, green
   and purple (from bottom to top)). We choose the $u_H=R=1$. } \label{fig12} 
\end{figure}

For simplicity, we mainly consider the case: $x_3'=0$, $\sin\theta=1$, $v=0$ and
a constant angular velocity $\omega$ in the following. It means the dipole is
vertical with the coordinate $x_3$, and the shape of string is independent to
the coordinate $x_3$. So the determinant of the
induced metric on string worldsheet becomes \be-g_2=-\det g_{\alpha\beta}=
g_{xx}(H-h\rho^2\omega^2)(g_{xx}h+g_{uu}u'^2).\label{determinant2}\ee Since it
must be non-negative, which implies there exists a constraint $1 \geq
K+h\rho^2\omega^2$. Thus, for a given angular velocity $\omega$, the variable
$u(\rho)$ has a bound. We explicitly show this bound in the figure
\ref{fig12}.
The string shape must be located above the corresponding curve for a fixed value $\omega$.

\begin{figure}[ht] 
\centering
 \includegraphics[width=0.55\textwidth]{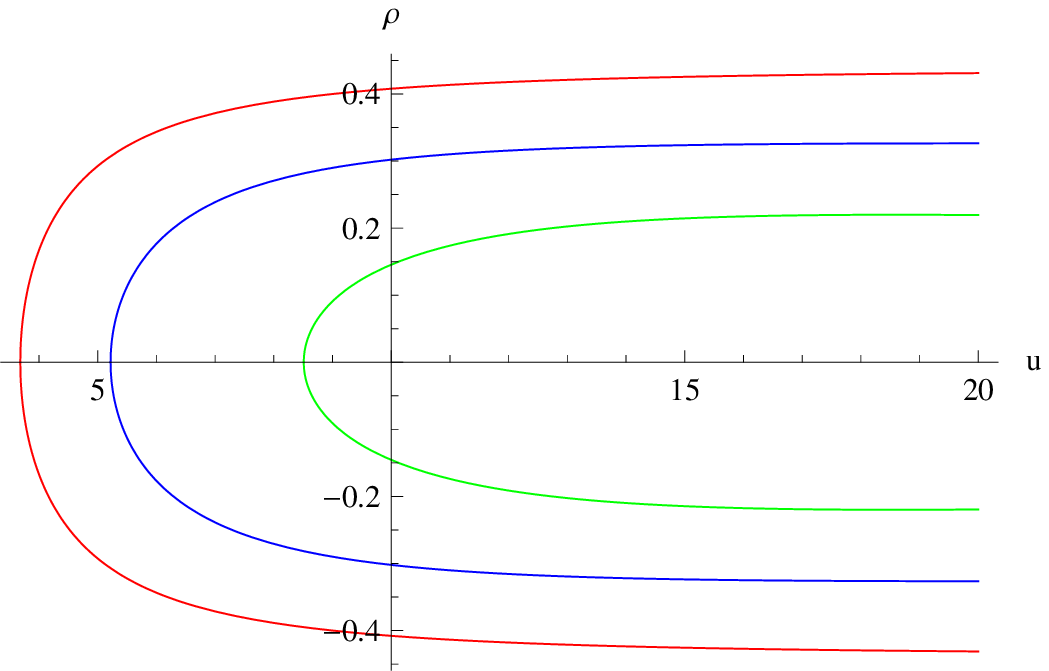}
\caption{It is the string shape with $\omega=1, ~1.5, ~3$ (red, blue and green) or (from left to 
 right) and $b=0$. }
\label{fig13}
 \begin{tabular}{cc}
 \includegraphics[width=0.46\textwidth]{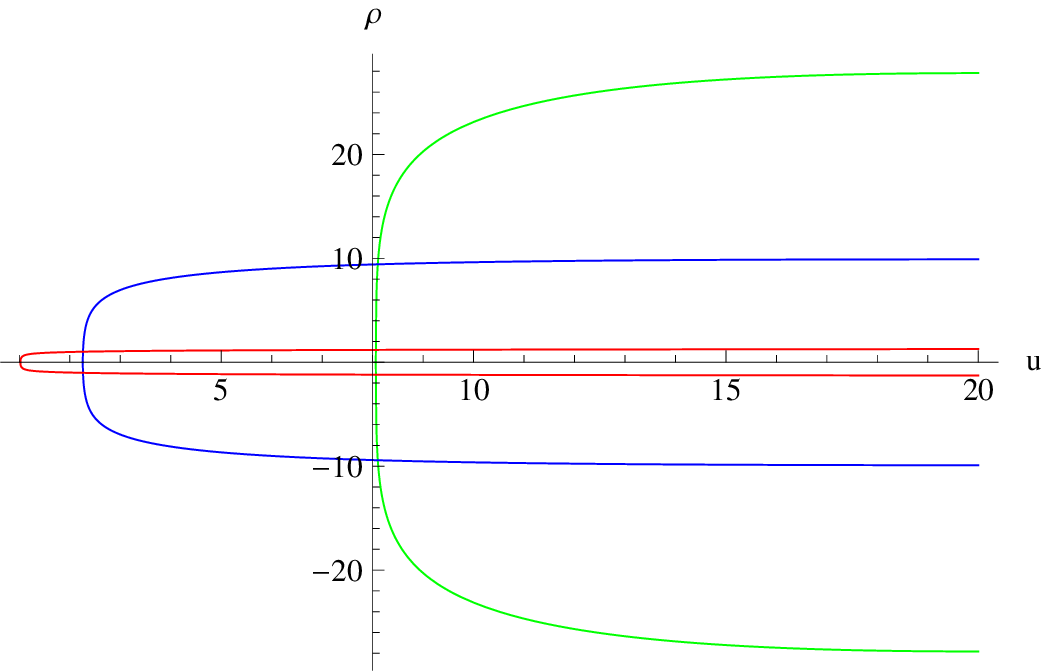}&
 \includegraphics[width=0.46\textwidth]{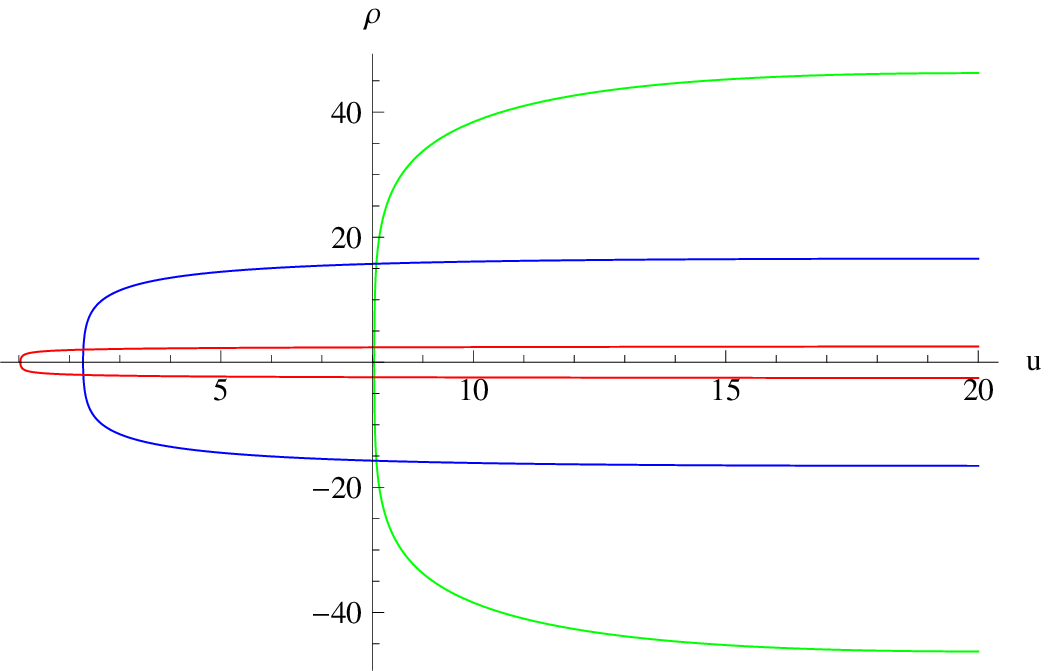}\\
   (a)   & (b) 
 \end{tabular}
 \caption{The shape of string is shown: (a) at $\omega=1,~1.5,~3$ (red, blue, green) or (from left to
 right) and $b=3$; (b) at $\omega=1,~1.5,~3$ (red, blue, green) or (from left to right) and $b=5$. }
 \label{fig14}
\end{figure}

Now the equations of motions
(\ref{equation1}) and (\ref{equation2}) become \bea
&&p=0 \nonumber \\
&&\partial_u(\sqrt{-g_2}-B_{12}\rho\omega)=\partial_\sigma\left({{(H-h\rho^2\omega^2)u'}\over{H\sqrt{-g_2}}}
\right). 
\label{equation4}\eea 
Then we get the shape of this rotating string shown in the figure
\ref{fig13} and \ref{fig14}. 
 Here we only plot the zero node (the times of crossing the $u$
coordinate) solution and already set $R=u_H=1$. With increasing the angular
velocity $\omega$ at $b=0$, the distance between quark and anti-quark is
became smaller, and the lowest point of this string is increased. But at a
finite $b$, the separation of quark and anti-quark is increased with increasing
the angular velocity $\omega$. At a fixed
$\omega$, the lowest point of this spinning sting is became smaller with
increasing the non-commutative parameter $b$.

With the equation (\ref{NG}) and (\ref{determinant2}), we obtain the energy
and the angular momentum for this spinning open string. The string energy is \be
E={{1}\over{\pi\alpha'}}\int_0^{{{L}\over{2}}} d\sigma
{{g_{xx}^2}\over{\sqrt{-g_2}}}\left((1-K)(h+{g_{uu}\over{g_{xx}}}u'^2)\right).\label{energy}\ee
If subtracting the self-energy of the isolated quark and anti-quark, one can get
the finite bound energy of the $q\bar{q}$. And angular momentum is derived as
\be J={{1}\over{\pi\alpha'}}\int_0^{{{L}\over{2}}} d\sigma
{{\omega\rho^2g_{xx}^2h}\over{\sqrt{-g_2}}}\left(h+{g_{uu}\over{g_{xx}}}u'^2\right)-
\frac{1}{\pi\alpha'}
\int_0^{\frac{L}{2}}d\sigma B_{12}\rho. \label{angularmomentum}\ee It
corresponds to the spin of a meson $q\bar{q}$ in the boundary theory. Using the
equations (\ref{energy}) and (\ref{angularmomentum}), the Regge trajectory
i.e. the relations between $E^2$ and $J$, can be derived.

\begin{figure}[ht] 
\begin{tabular}{cc}
 \includegraphics[width=0.46\textwidth]{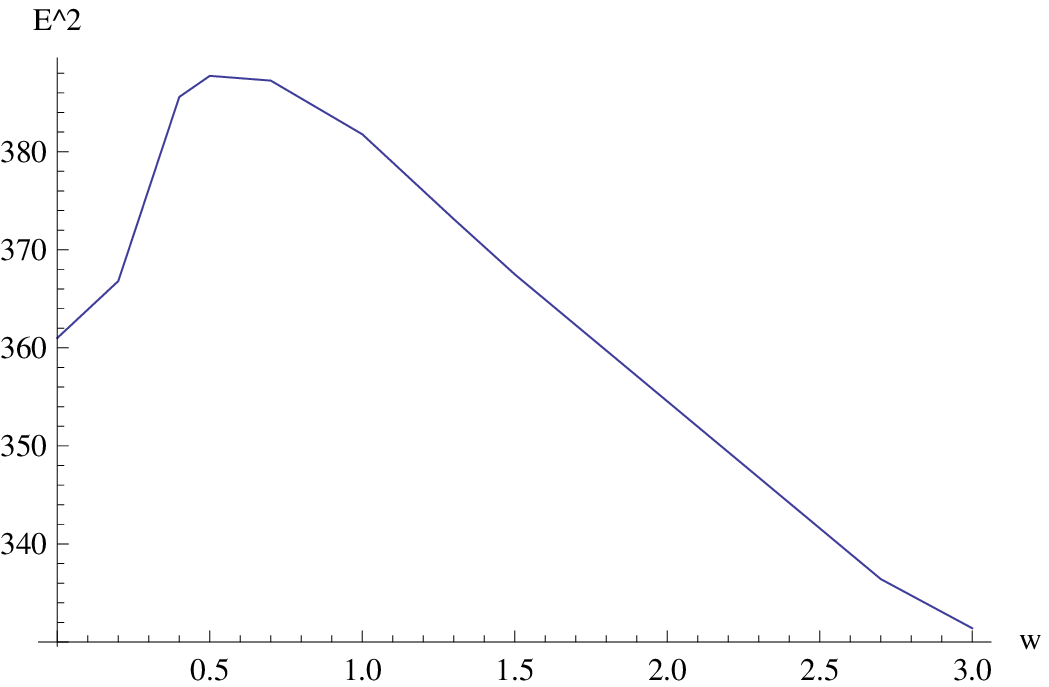}&
\includegraphics[width=0.46\textwidth]{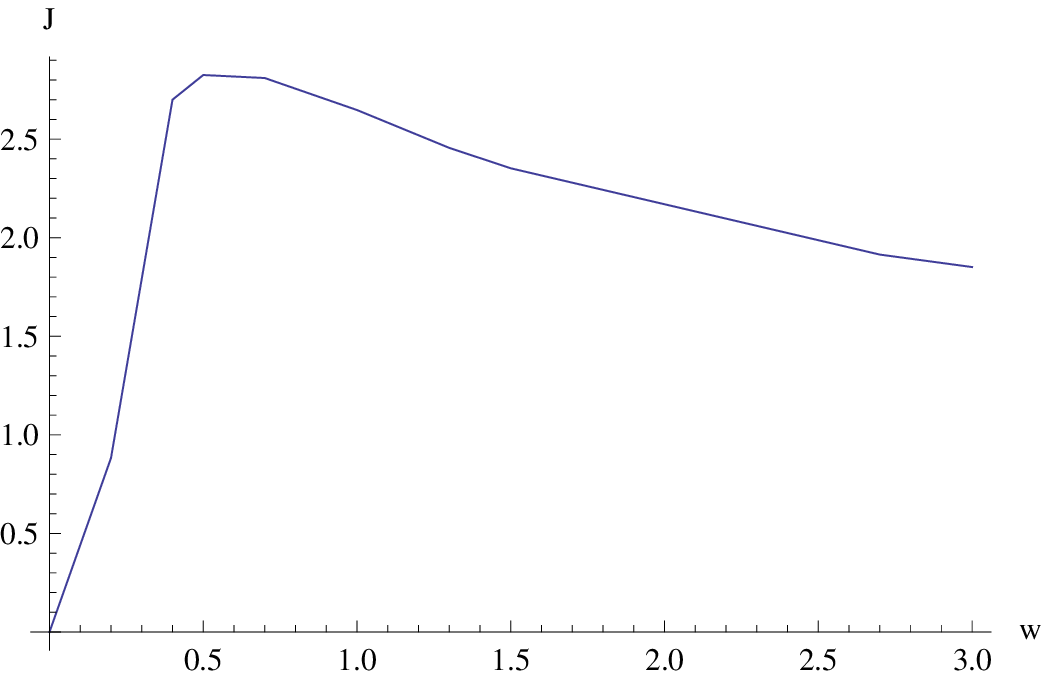}\\
   (a)   & (b) 
 \end{tabular}
\caption{(a) $E^2$ varies with the angular velocity $\omega$ at $b=0$; (b) The
  angular momentum $J$ varies with $\omega$ at $b=0$.  } 
\label{fig15}
\centering
\includegraphics[width=0.55\textwidth]{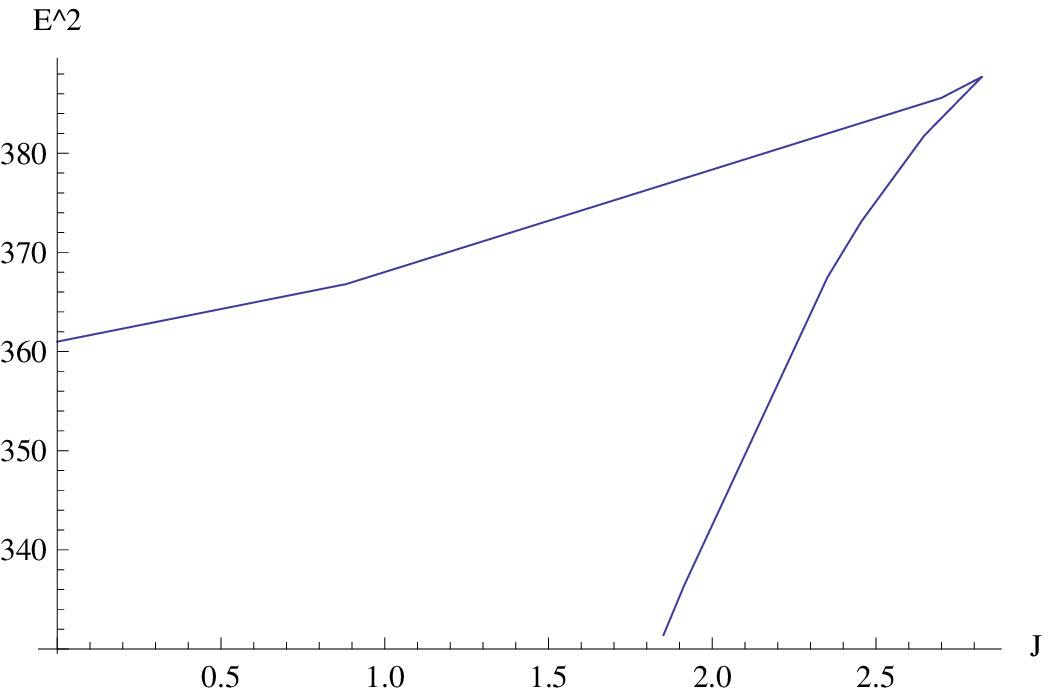}
 \caption{It is the Regge behavior $E^2(J)$ vs $J$ at $b=0$. }
 \label{fig16} 
\end{figure}

\begin{figure}[ht] 
\begin{tabular}{cc}
 \includegraphics[width=0.46\textwidth]{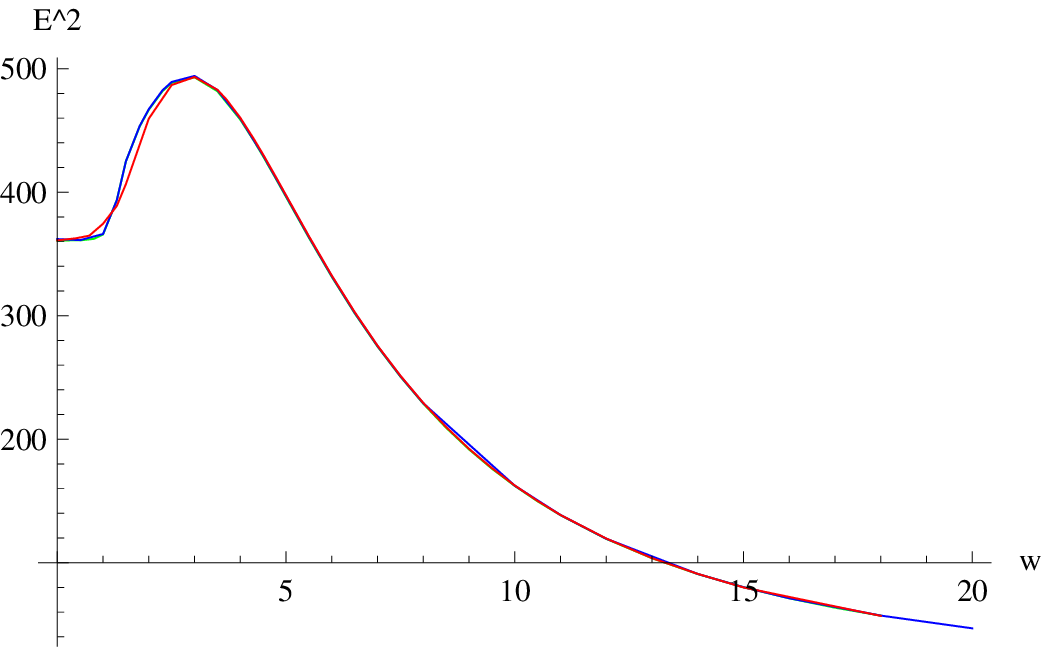}&
 \includegraphics[width=0.46\textwidth]{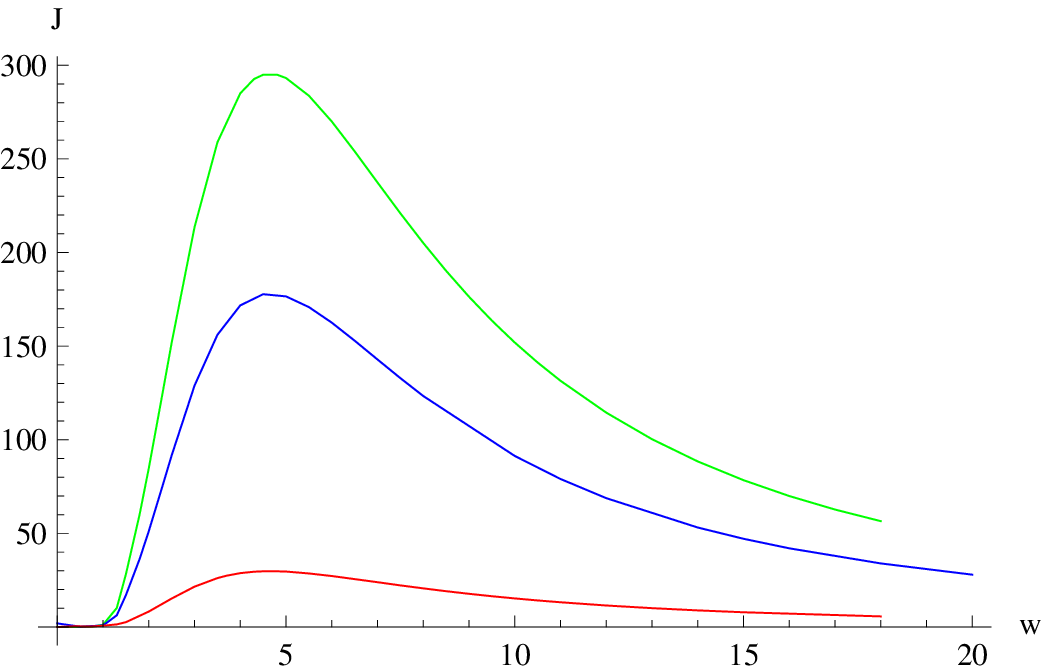}\\
   (a)   & (b) 
 \end{tabular}
\caption{(a) $E^2$ varies whit $\omega$ at $b=0.5,~3,~5$ (red, blue, green). 
The curved lines $E^2(\omega)$ is almost same at different constant value $b$; 
(b) It is shown the relation between $J$ and $\omega$ at $b=0.5,~3,~5$ (red, blue, green) 
or (from bottom to top).}
\label{fig17}
 \centering
 \includegraphics[width=0.55\textwidth]{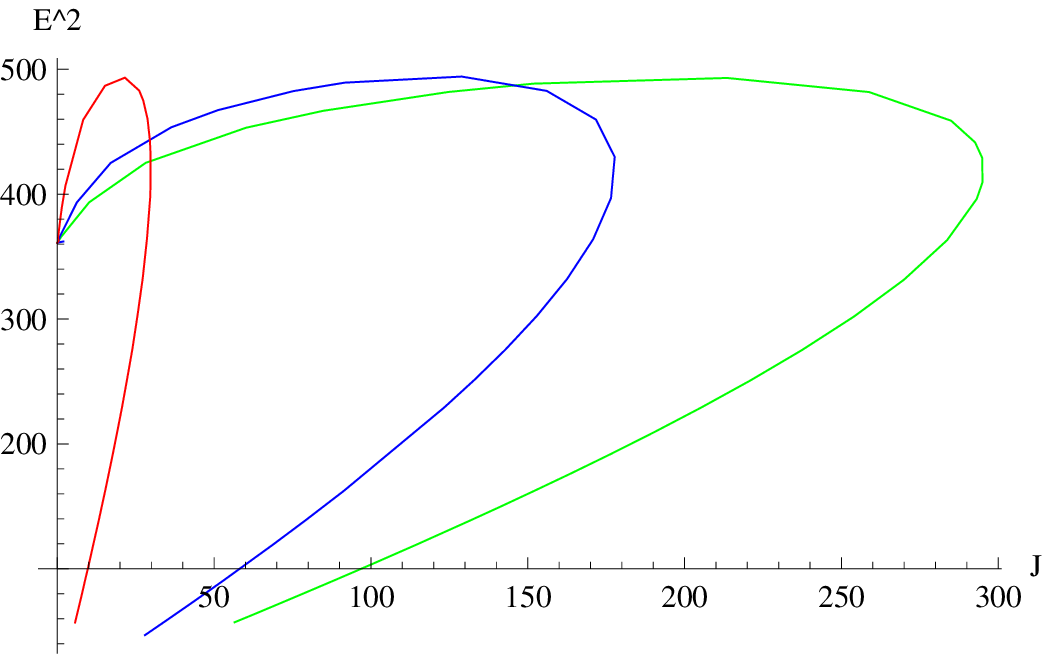}
 \caption{$E^2$ varies with the angular momentum $J$ at $b=0.5,~3,~5$ (red,
   blue, green) or (from left to right). }
 \label{fig18} 
\end{figure}

\begin{figure}[ht] 
\begin{tabular}{cc}
 \includegraphics[width=0.46\textwidth]{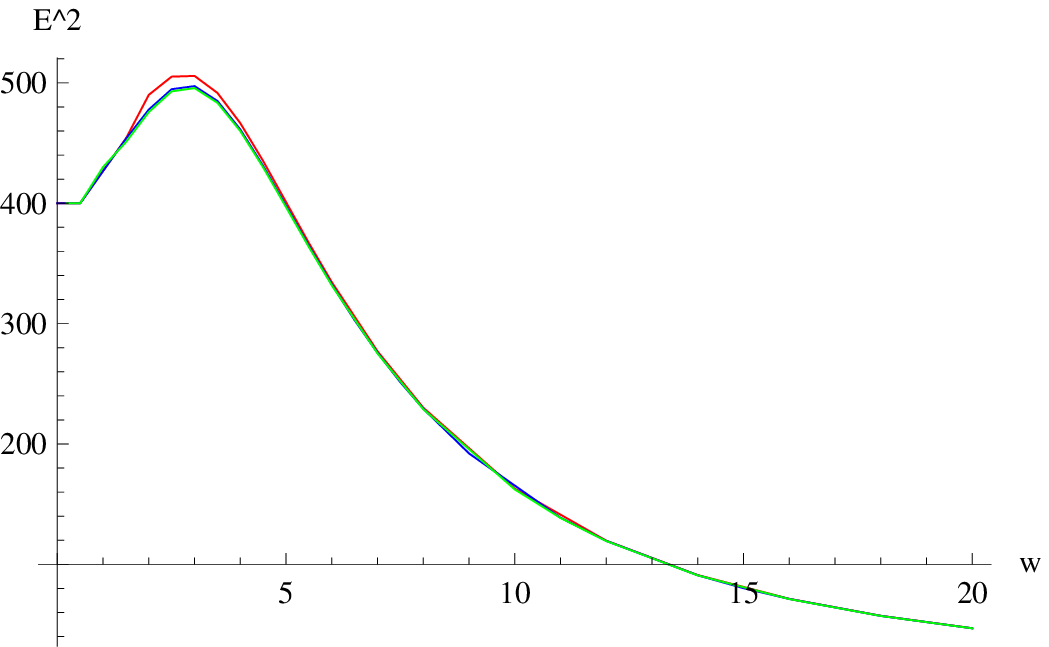}&
 \includegraphics[width=0.46\textwidth]{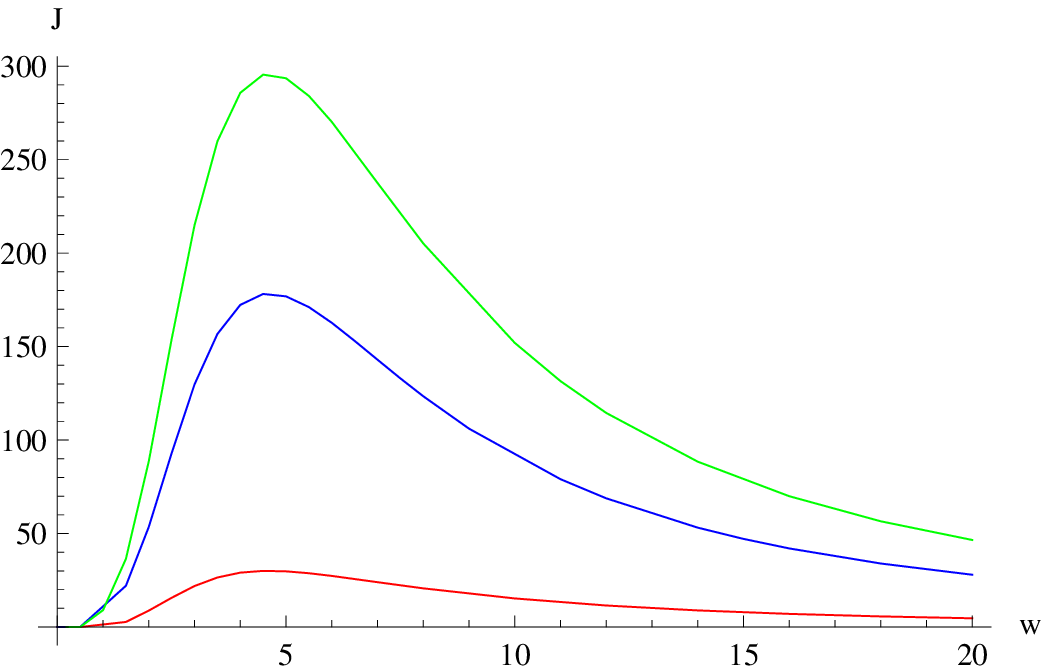}\\
   (a)   & (b) 
 \end{tabular}
\caption{(a) $E^2$ varies with $\omega$ at $b=0.5,~3,~5$ (red, blue, green) and
  $T=0$. These three curves are almost overlaped each other; (b) $J$ varies with $\omega$ at
  $b=0.5,~3,~5$ (red, blue, green) or (from bottom to top) and $T=0$. } 
\label{fig19}
 \centering
 \includegraphics[width=0.55\textwidth]{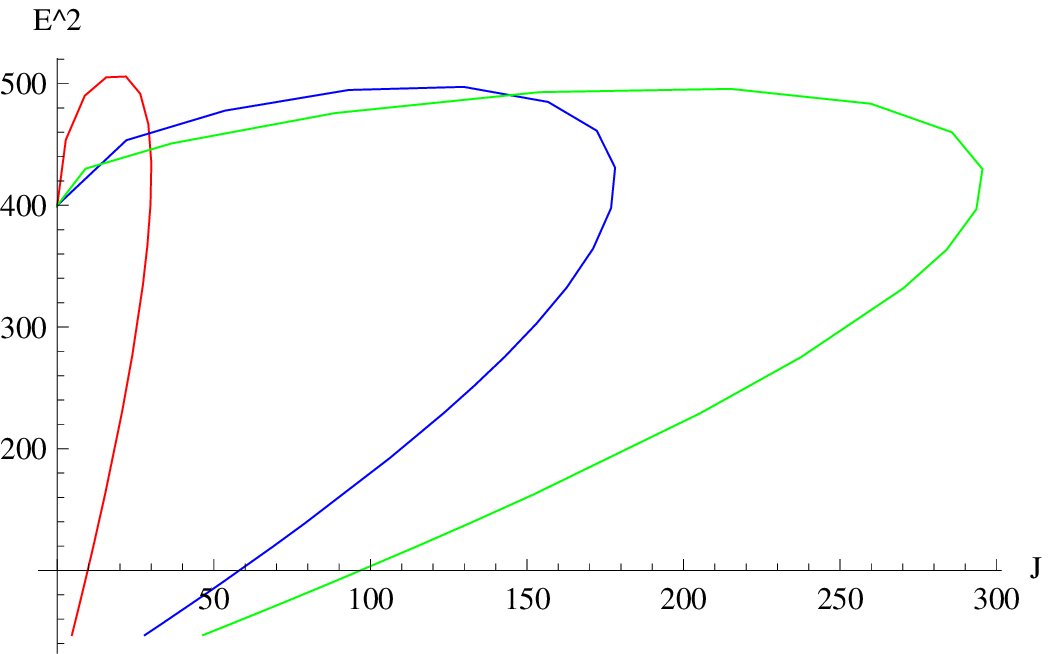}
 \caption{This is the figure about $E^2$ varying with $J$ at $b=0.5,~3,~5$ (red,
   blue, green) or (from left to right) and $T=0$. } 
 \label{fig20} 
\end{figure}

Now we investigate the relations $E^2(\omega)$, $J(\omega)$ and
$E^2(J)$.  Firstly, we consider the $b=0$ case, and plot the figures
\ref{fig15} and \ref{fig16} in below\footnote{ we set $R=u_H=1$ and
  $u_{\infty}=20$ in the figures \ref{fig15}-\ref{fig20}}. The
energy square $E^2$ and the angular momentum $J$ have a maximum value at a special angular
velocity value $w$. The function $E^2(J)$ has two branches relative to the
angular momentum $J$. With increasing the
angular momentum (increasing the angular velocity), the $E^2$ is
increased. After a maximum value, the value $E^2$ is decreased with decreasing
the angular momentum $J$ (but increasing the angular velocity). Since our model
reduces to the Sakai-Sugimoto model at the case $b=0$, these results here is same as
in \cite{Johnson:2009ev} and \cite{Peeters:2006iu}.

Then we consider a finite non-commutative parameter $b$ at finite temperature. The
figures \ref{fig17} and \ref{fig18} are plotted. Explicitly, the curves in
figures still have these properties as in \ref{fig15} and
\ref{fig16}. However, because of the effects of the NS-NS $B_{12}$ field, the curves
become more smoothly. With increasing the parameter $b$, the energy square $E^2$
almost isn't changed, but the angular momentum $J$ is increased.

If the temperature vanishes, i.e. $u_H=0$, and $b$ is a finite value, then the
behaviors of the $E^2(\omega)$, $J(\omega)$ and $E^2(J)$ are shown in the
figures \ref{fig19} and \ref{fig20}. These results are very similar to the
finite temperature cases.
And if let the temperature and parameter to vanish together, then such results
are already investigated in \cite{Johnson:2009ev} and
\cite{Peeters:2006iu}. From the figures \ref{fig15}-\ref{fig20}, it is clear
when the angular velocity $\omega$ is equal to zero, then the angular momentum $J$ vanishes.

If chosen the angle between the dipole and $x_3$ is $\theta=0$, then the wind is
parallel to the q-${\rm\bar{q}}$ dipole. Now the worldsheet coordinates parameterization
(\ref{parameterization1}) is not well defined. Instead, the string worldsheet
gauge is chosen as \be ~\tau=t, ~~~\sigma=x_3, ~~~\varphi=\omega t, ~~~
u(\sigma),~ ~~\rho(\sigma). \ee Then the induced metric, its determinant and the
pullback of NS-NS $B_{12}$ field are listed as \bea &&g_{\tau\tau}= -A+g_{xx}h\rho^2\omega^2,
~~g_{\tau\rho}=B,~~
g_{\sigma\sigma}=C+g_{xx}h\rho'^2+g_{uu}u'^2, \nonumber \\
&&-g_3=g_{xx}^2\left(1-K-(1+K\sinh^2\eta)h\rho^2\omega^2\right. \cr
&&~~~~~~~~~~~~
\left. +(1-K\cosh^2\eta-h\rho^2\omega^2)(h\rho'^2+{{g_{uu}}\over{g_{xx}}}u'^2)\right),\cr
&& P[B]_{\tau\sigma}=B_{12}\rho\rho'\omega.\eea Since the Lagrangian does not
explicitly depend on the variable $\sigma$, the Hamiltonian is conserved with
respect to the variable $\sigma$. Then the conserved constant $q$ is defined as
\be q =
{{g_{xx}^2}[1-K-(1+K\sinh^2\eta)\rho^2\omega^2]\over{\sqrt{-g_3}}}.\label{conserved}\ee
And the other equations of motion are \bea
&&\partial_\sigma\left({{g_{xx}^2(1-K\cosh^2\eta-h\rho^2\omega^2)h\rho'}\over{\sqrt{-g_3}}}-B_{12}
  \rho\omega\right) =\cr &&\quad\quad{{\rho
    g_{xx}^2h\omega^2(1+K\sinh^2\eta+h\rho'^2+
    {{g_{uu}}\over{g_{xx}}}u'^2)}\over{\sqrt{-g_3}}}-B_{12}\omega\rho',
\nonumber\\
&& \partial_u(\sqrt{-g_3}-B_{12}\omega\rho\rho')=
\partial_\sigma\left({{g_{xx}g_{uu}(1-K\cosh^2\eta-h\rho^2\omega^2)u'}\over
    {\sqrt{-g_3}}}\right).\label{equation3}\eea One can use the equation
(\ref{conserved}) and the equations of motion (\ref{equation3}) to
calculate the dynamics of a open string in the gravity background (\ref{metric6}). In addition, the
energy and angular momentum of open spinning string at the $\theta=0$ case are \bea
E&=&{{1}\over{\pi\alpha'}}\int_0^{{L}\over{2}}d\sigma{{g_{xx}^2
    \left(1-K+(1-K\cosh^2\eta)(h\rho'^2+{{g_{uu}}\over{g_{xx}}}u'^2)\right)}\over{\sqrt{-g_3}}},
\nonumber \\
J&=&{{1}\over{\pi\alpha'}}\int_0^{{L}\over{2}}d\sigma{{\omega\rho^2g_{xx}^2h
    \left(1+K\sinh^2\eta+h\rho'^2+{{g_{uu}}\over{g_{xx}}}u'^2\right)}\over{\sqrt{-g_3}}}\cr
&& \quad\quad -{{1}\over{\pi\alpha'}}\int_0^{{L}\over{2}}d\sigma
B_{12}\rho\rho'.
\label{angular} \eea If the coordinate $\rho$ is constant, and $\omega=0$, then
it reduces to the similar case studied in \cite{Liu:2006nn}-\cite{Caceres:2006ta}. Otherwise, if
$v=0$, then the metric determinant and the equations (\ref{conserved}),
(\ref{equation3}) and (\ref{angular}) reduce to \bea
&&-g_4=g_{xx}^2(1-K-h\rho^2\omega^2)(1+h\rho'^2+{{g_{uu}}\over{g_{xx}}}u'^2),\cr
&&q = {{g_{xx}^2}(1-K-h\rho^2\omega^2)\over{\sqrt{-g_4}}},\cr
&&\partial_\sigma\left({{g_{xx}^2h(1-K-h\rho^2\omega^2)\rho'}\over{
      \sqrt{-g_4}}}-B_{12}\rho\omega\right) = \cr &&\quad\quad {{\rho
    g_{xx}^2h\omega^2(1+h\rho'^2+{{g_{uu}}\over{g_{xx}}}u'^2)}\over{
    \sqrt{-g}}}-B_{12}\omega\rho', \cr
&&\partial_u(\sqrt{-g_4}-B_{12}\omega\rho\rho')=
\partial_\sigma\left({{g_{xx}g_{uu}(1-K-h\rho^2\omega^2)u'}\over{\sqrt{-g_4}}}
\right),\\
&&E={{1}\over{\pi\alpha'}}\int_0^{{L}\over{2}}d\sigma{{g_{xx}^2(1-K)(1+h\rho'^2+{{g_{uu}}
      \over{g_{xx}}}u'^2)}\over{\sqrt{-g_4}}},\cr
&&J={{1}\over{\pi\alpha'}}\int_0^{{L}\over{2}}d\sigma{{\omega\rho^2g_{xx}^2h
    \left(1+h\rho'^2+{{g_{uu}}\over{g_{xx}}}u'^2\right)}\over{\sqrt{-g_4}}}-{{1}\over{\pi\alpha'}}
\int_0^{{L}\over{2}}d\sigma B_{12}\rho\rho'.\nn\eea One can discuss the
relations of the energy $E$, angular momentum $J$ with the angular velocity
$\omega$, and investigate the Regge trajectory behavior $E^2(J)$. Here we only
list these equations, the numerical calculations can be done by using the same
method as before.

\section{Summaries}
In this paper, we construct a holographic model by the
D4-D8/${\rm\overline{D8}}$ brane configuration with a NS-NS $B_{12}$ field. The
effective theory on the intersecting region of this brane configuration is a
four-dimensional non-commutative field theory. In the strong coupling regime, we
investigate some underlying low energy dynamics of this theory by using the
supergravity/Born-Infeld approximation.

If don't turn on the gauge field on the flavor D8-brane, we find the effective
D8-brane action is similar to the commutative case \cite{Sakai:2004cn} except for
some coefficients. So all
the dynamics, for example the chiral symmetry breaking, are same. However, with
a magnetic field along the direction $x_1$ and $x_2$, then the DBI 
action of the flavor D8-brane includes the non-commutative parameter $b$ and
magnetic field parameter $B$. We analyze their influences on the asymptotic
distance $L$ between D8 and ${\rm\overline{D8}}$ brane and the energy difference
$\delta S$ of two solutions. In both the low temperature and high temperature
phase, the distance $L$ is increased with increasing the magnetic parameter at a
fixed value $b$. However, the thing is conversed if increasing the
non-commutative parameter $b$ at a fixed value $B$. It means the parameter $B$
and $b$ have converse contributions to the asymptotic distance $L$. For the
energy difference, it is always negative with arbitrary value $B$ and $b$ at low
temperature. So the chiral symmetry is independent on the magnetic field and
non-commutative parameter. And it is always broken. In the high temperature
phase, the energy difference $\delta S$ between the connected and separated
solutions has a critical temperature $T_\chi$. At $b=0$, our model reduces to
the Sakai-Sugimoto model with a magnetic field, and here the results is same as
in \cite{Johnson:2008vna}. At a finite fixed value $b$, the critical point $y_H$
becomes smaller with increasing $B$. However, it become larger with increasing
$b$ at a fixed $B$. As similar as the distance $L$, the contributions of the $B$
and $b$ are also conversed. This may be understood from the equation
(\ref{NSNS}), in which the parameter $B$ is proportional to $1/b$. Below the
critical points, the energy difference $\delta S$ is negative. Then the
connected D8-${\rm\overline {D8}}$ brane solution is dominated, the chiral
symmetry is broken $U(N_f)\times U(N_f)\rightarrow U(N_f)_{\rm
  diag}$. Otherwise, $\delta S$ is positive, and the dominated solution is the
separated D8-${\rm\overline{D8}}$ configuration, and the chiral symmetry will be
restored. Thus there exists a first order chiral phase transition at these
critical points.

We also generalize to consider some other branes configurations. Some three- and
four-dimensional non-commutative field theories are constructed through some
intersecting brane configurations. The results is similar to the cases of
D8-${\rm\overline{D8}}$ brane configurations. Maybe these models have some
applications to study the condensed matter physics or something else.

And we investigated the dynamics of a fundamental string in the high temperature
background (\ref{metric6}). We find the non-commutative parameter will decrease
the drag force. The relations $E^2(\omega)$, $J(\omega)$ and $E^2(J)$ become
more smoothly than the results without the non-commutative effects. With
increasing the parameter $b$, the energy $E^2$ is almost unchanged, and the
angular momentum $J$ is became larger.

There are some generalizations to this holographic model. The first one is to
study the meson spectra through investigating the world-volume field
fluctuations on the flavor brane. The fluctuation DBI action is similar to the
commutative case just like in \cite{Arean:2005ar}, however, the Chern-simons
term contains some non-commutative contributions. Then one can investigate the
Nambu-Goldstone bosons corresponding to the chiral symmetry breaking in the
meson spectra. The second is to consider a chemical potential in this and other
general holographic models by using the method in
\cite{Horigome:2006xu}. Also one can calculate the shear
and bulk viscosity of non-commutative QGP as the similar way in
\cite{Landsteiner:2007bd}. Finally, following in
\cite{Hata:2007mb}, to construct the baryons in this
holographic model is also interesting.

\subsection*{Acknowledgments} 
Wei-shui thank CQUeST for providing an office. It is more convenient for me to
study and research. This work of Yunseok was supported by the Korea Science and
Engineering Foundation (KOSEF) grant funded by the Korea government(MEST)
through the Center for Quantum Spacetime(CQUeST) of Sogang University with grant
number R11-2005-021. This work of SJS and Weishui is supported by KOSEF Grant
R01-2007-000-10214-0 and also by the SRC Program of the KOSEF through the Center
for Quantum SpaceTime (CQUeST) of Sogang University with grant number
R11-2005-021.

\appendix
\section{D-brane backgrounds with a NS-NS  field}
In this appendix, we give the gravity backgrounds used in the main part of this
paper.

Turning on a constant NS-NS $B_{MN}$ background field in a flat spactime, the open
string in this background can be quantized. Then, due to this $B_{MN}$ field, the
field theory on the D-brane worldvolume is non-commutative
\cite{Seiberg:1999vs}. The non-commutative parameter is proportional to this
constant NS-NS $B_{MN}$ field.

The gravity solution of D-brane with a NS-NS $B_{MN}$ field was constructed in
\cite{Breckenridge:1996tt}-\cite{Alishahiha:1999ci}. In the
following, we mainly focus on the D4-brane backgrounds with a NS-NS $B_{MN}$
background field. This
supergravity solution is \bea
&&ds^2 = f^{-1/2}\left[-dt^2+h(dx_1^2+dx_2^2)+dx_3^2+dx_4^2\right]+f^{1/2}(dr^2+r^2d\Omega_4^2),\cr
&& f=1+\frac{\alpha' R^3}{r^3},\quad R^3=\frac{\pi^2g_{\rm
    YM}^2N_c}{4\cos\theta},\cr
&& h^{-1}=\sin^2\theta f^{-1}+\cos^2\theta, \label{metric1} \\
&& B_{1 2}=hf^{-1}\tan\theta, \quad e^{2\phi}=g_s^2f^{-1/2}h,\nn \\
&& C_{01234}=g_s^{-1}(1-f^{-1})h\cos\theta, \quad
C_{034}=g_s^{-1}(1-f^{-1})\sin\theta.\nn \eea where $g_{\rm
  YM}^2=(2\pi)^2g_sl_s$. Actually, it is a bound state of the D2 and D4 brane
\cite{Breckenridge:1996tt}. The D2-brane extends along the $t,~ x_3$ and $x_4$
directions, and D4-brane's worldvolume coordinates are $t$ and $x_{1,\cdots,
  4}$. This solution can be got by applying the delocalization and rotation on a
D3-brane and doing T-duality. The NS-NS $B_{12}$ field is produced through the
T-duality acting on the non-trivial gravity background.

Define some parameters as \be u=r/\alpha', ~~~ g_s'=\frac{g_sb}{l_s}, ~~~
b=\alpha'\tan\theta,~~~ x_{1,2}'=\frac{b}{\alpha'}x_{1,2}. \label{decoupling}\ee
If choose the limit $\alpha'\rightarrow 0$, and let the above parameters to be
fixed, then the gravity solution (\ref{metric1}) becomes\footnote{We already
  replace the $g_s', x_{1,2}'$ by $g_s, x_{1,2}$, and will set $\alpha'=1$ in
  the main part of this paper. } \bea
&&ds^2=\left(\frac{\alpha'^2u^3}{R^3}\right)^{1/2}\left[-dt^2+h(dx_1'^2+dx_2'^2)+dx_3^2+dx_4^2\right]
+\left(\frac{R^3}{\alpha'^2u^3}\right)^{1/2}(du^2+u^2d\Omega_4^2),\cr
&&R^3=\pi^4g_s' N_c, \quad h=\frac{1}{1+a^3u^3},\quad
a^3\equiv\frac{b^2}{R^3}, \nn \\
&& B_{12}=\frac{\alpha'a^3u^3}{b(1+a^3u^3)}=\frac{\alpha'}{b}(1-h), \quad
e^{2\phi}=g_s'^2\frac{\alpha'^2}{R^{3/2}b^2}u^{3/2}h, \label{metric2}\\
&& C_{01234}=g_s'^{-1}\alpha'^{1/2}h,\quad C_{034}=g_s'^{-1}\alpha'^{-1/2}b.\nn
\eea The gravity theory on this near horizon background is dual to a
five-dimensional non-commutative field theory. Here the coordinates $x_1$ and $x_2$
in the boundary theory is non-commutative, $[x_1, x_2]\sim b$. One can
generalize to consider some other bound states. For example, if consider the
D4-D2-D2-D0 case, then one find the space coordinate $x_3$ and $x_4$ of the boundary
field theory will be non-commutative \cite{Breckenridge:1996tt} and \cite{Alishahiha:1999ci}.

From the solution (\ref{metric2}), the curvature scalar is \be \mathcal{R}\sim
\frac{1}{g_{eff}},\quad g_{eff}\sim g_sN_cu.\ee And the effective string
coupling constant is \be e^\phi\sim \frac{g_{eff}^{3/2}}{N_c\sqrt{1+a_{eff}^3}},
\quad ~ a_{eff} \sim \left(\frac{bu^2}{g_{eff}}\right)^{2/3},\ee where $a_{eff}$
is an effective non-commutative parameter. if $a_{eff}\ll 1$, then it means the
non-commutative effect can be neglected, and the non-commutative field theory
will reduce to a commutative theory. when the following condition \be
\mathcal{R}\ll 1, \quad e^\phi\ll 1\ee are satisfied, the low energy dynamics on
the D-brane can be investigated by the supergravity approximation. More detail
discussions can be found in \cite{Alishahiha:1999ci}.

A non-extremal generalization of the supergravity solution (\ref{metric1}) is
\bea
&&ds^2=f^{-1/2}\left[-Hdt^2+h(dx_1^2+dx_2^2)+dx_3^2+dx_4^2\right]+f^{1/2}(H^{-1}dr^2+r^2d\Omega_4^2),
\nn\\
&& H=1-\frac{r_H^3}{r^3}, \quad R^3=\frac{\pi^2g_{\rm
    YM}^2N_c}{4\cos\theta},\cr
&& h^{-1}= \sin^2\theta f^{-1}+\cos^2\theta, \label{metric3} \\
&& B_{1 2}=hf^{-1}\tan\theta, \quad e^{2\phi}=g_s^2f^{-1/2}h,\nn \\
&& C_{01234}=g_s^{-1}(1-f^{-1})h\cos\theta, \quad
C_{034}=g_s^{-1}(1-f^{-1})\sin\theta.\nn \eea Then, in the decoupling limit
(\ref{decoupling}), the near horizon geometry of the metric (\ref{metric3}) is
\bea
&&ds^2=\left(\frac{u}{R}\right)^{3/2}\left[-Hdt^2+h(dx_1^2+dx_2^2)+dx_3^2+dx_4^2\right]+
\left(\frac{R}{u}\right)^{3/2}(H^{-1}du^2+u^2d\Omega_4^2),
\nn \\
&& H=1-\frac{u_H^3}{u^3}.\label{metric4} \eea The other background fields are
same as the solution (\ref{metric2}). The corresponding effective theory in the
boundary of the gravity background (\ref{metric4}) is a finite temperature
generalization to the zero temperature case of the gravity background (\ref{metric2}).

\end{document}